\documentclass[letterpaper,twocolumn,10pt]{article}
\usepackage{multirow}
\usepackage{usenix-2020-09}

\usepackage{tikz}
\usepackage{pgfplots}
\pgfplotsset{compat=1.18}
\usepackage{amsmath}
\usepackage{amssymb}
\usepackage{xspace}
\usepackage{xcolor}
\usepackage{algorithm}
\usepackage{algorithmicx}
\usepackage{algpseudocode}
\usepackage{listings}
\usepackage{booktabs}
\usepackage{graphicx}
\usepackage{siunitx}  \usepackage[available]{usenixbadges}

\microtypecontext{spacing=nonfrench}
\newcommand{\sysname}{Tady\xspace}

\begin{document}

\date{}

\title{\Large \bf \sysname : A Neural Disassembler without Structural Constraint Violations}

\author{{\rm Siliang Qin$^{1,2}$, Fengrui Yang$^{3}$, Hao Wang$^{3}$, Bolun Zhang$^{1,2}$, Zeyu Gao~$^{3}$, Chao Zhang~$^{3}$\thanks{Corresponding authors.}, Kai Chen~$^{1,2}$\addtocounter{footnote}{-1}\footnotemark}\\
\normalsize\textit{$^1$ Institute of Information Engineering, Chinese Academy of Sciences, China}\\
\normalsize$^2$\textit{School of Cyber Security, University of Chinese Academy of Sciences, China}\\
\normalsize$^{3}$\textit{Tsinghua University, China}\\
\textit{\{qinsiliang, zhangbolun, chenkai\}@iie.ac.cn} \\
\textit{\{yangfr23, hao-wang20, gaozy22\}@mails.tsinghua.edu.cn, chaoz@tsinghua.edu.cn}
}

\maketitle

\begin{abstract}
Disassembly is a crucial yet challenging step in binary analysis. While emerging neural disassemblers show promise for efficiency and accuracy, they frequently generate outputs violating fundamental structural constraints, which significantly compromise their practical usability.
To address this critical problem, we regularize the disassembly solution space by formalizing and applying key structural constraints based on post-dominance relations. This approach systematically detects widespread errors in existing neural disassemblers' outputs. These errors often originate from models' limited context modeling and instruction-level decoding that neglect global structural integrity. We introduce \sysname, a novel neural disassembler featuring an improved model architecture and a dedicated post-processing algorithm, specifically engineered to address these deficiencies.
Comprehensive evaluations on diverse binaries demonstrate that \sysname effectively eliminates structural constraint violations and functions with high efficiency, while maintaining instruction-level accuracy. 

\end{abstract}

\section{Introduction}
\label{sec:introduction}

Disassembly, the process of identifying instructions from raw byte sequences, forms the foundation of binary analysis~\cite{Popa2012BinaryCD, khadraSpeculativeDisassemblyBinary2017, Pang2020SoKAY}. Its accuracy is paramount, as numerous subsequent sophisticated analyses critically depend on its output. These include decompilation~\cite{Sisco2023LoopRF, cifuentesDecompilationBinaryPrograms1995, Liang2021NeutronAA}, data flow analysis~\cite{Zhiyong2010MethodBO, Yongcheng2010ProgramUA}, program slicing~\cite{cifuentesIntraproceduralStaticSlicing1997a}, and binary code similarity detection~\cite{Qiao2021MultiLevelCB, huSemanticsBasedHybridApproach2021, Wang2024CLAPLT, Wang2024CEBinAC}. Errors introduced during disassembly can cascade, potentially invalidating entire analytical efforts.

Traditional disassembly algorithms contend with fundamental ambiguities inherent in binary code. These challenges include distinguishing code from data within the same memory space and precisely identifying instruction boundaries, especially in architectures featuring variable-length instructions. Industry-standard disassemblers such as IDA Pro, Ghidra, and Binary Ninja~\cite{hex-raysIDAPro2025, nationalsecurityagencyGhidra2025, vector35inc.BinaryNinja2025} employ sophisticated heuristics~\cite{Zhoujun2011DisassemblyMB} to navigate these complexities. While often effective, their reliance on heuristics limits their adaptability and can lead to failures in corner cases~\cite{flores-montoyaDatalogDisassembly2020a,Yu2022DeepDiLA,basqueAhoySAILRThere2024} or obfuscations~\cite{Linn2003ObfuscationOE,Popov2007BinaryOU}. 

Recent advancements in learning-based disassemblers~\cite{Yu2022DeepDiLA, Pei2020XDAAR, Wartell2011DifferentiatingCF} offer a promising data-driven paradigm, reducing dependence on handcrafted heuristics. These models, trained on extensive datasets of labeled binaries, first perform superset disassembly, which assumes all addresses as potential instruction starts and then categorizes them to find true code. They achieve high accuracy at the individual instruction level. However, a significant hurdle impedes their widespread adoption: \textbf{violation of fundamental constraints}. For example, a model might incorrectly identify a fall-through byte sequence after a valid instruction as non-code, leading to an output that represents an invalid execution trace. These violations render the disassembly practically unusable for downstream analyses, despite potentially high instruction-level metrics.

While individual instruction predictions might be locally accurate, a valid disassembly necessitates that choices for different byte sequences are consistent with each other. These disassembly choices exhibit complex interdependence, which can be broadly categorized into: \textit{mutual exclusions}, where one disassembly choice invalidates another (e.g., overlapping instruction candidates cannot both be valid code), and \textit{structural implications}, where the validity of one instruction choice necessitates or is predicated upon the validity of another due to inherent program structure (e.g., relationships dictated by control flow). Effectively modeling and enforcing the interdependence can significantly \textbf{regularize the solution space} of the inherently ill-posed disassembly problem, thereby enhancing the reliability of the results. Building on this principle, we present \textit{\sysname}, a novel approach designed to overcome key technical challenges in leveraging these structural constraints.

A primary challenge lies in \textbf{systematically characterizing and efficiently detecting violations of inter-instruction constraints.} While the concept of using interdependence to regularize disassembly is not new, prior approaches like Pdisasm~\cite{Miller2019ProbabilisticD} and D-Arm~\cite{Ye2023DARMDA} typically relied on formulating explicit pairwise constraints between instruction candidates. Although capable of capturing local relationships, this approach scales quadratically, becoming computationally intractable when applied to superset disassembly where the number of potential instructions can easily reach millions.

Our key insight is that the complex web of both mutual exclusion and structural implication constraints can be systematically and efficiently captured within the framework of \textbf{post-dominance relations} on a superset Control Flow Graph~(CFG). In compiler theory, an instruction $A$ is said to post-dominate an instruction $B$ if every path from $B$ to any program exit point must pass through $A$. This structural property is pivotal: for instance, if $B$ is determined to be code, then $A$ must also be code, otherwise, executing $B$ would lead to executing data at $A$; conversely, if $A$ is not valid code, then $B$ cannot be code, as this would imply an execution path that eventually attempts to execute an invalid instruction at $A$.

Building upon this insight, \textit{\sysname} characterizes these global constraints based on post-dominance. It leverages the post-dominator tree (PDT, introduced in Section~\ref{sec:PDT}) derived from the superset CFG to detect violations of these structural constraints. This method allows for an efficient, linear-time algorithm to identify such inconsistencies, thereby avoiding the prohibitive cost of exhaustive pairwise comparisons.

Another significant challenge is \textbf{representing long-range interdependence and execution-aware context.} To satisfy mutual exclusion and structural implication constraints, a disassembler must understand relationships between instructions that can be distant, both in memory addresses and execution paths. Existing learning-based approaches often prove insufficient. Graph-based models like DeepDi~\cite{Yu2022DeepDiLA}, despite using message passing, typically have a limited receptive field. Sequence-based models like XDA~\cite{Pei2020XDAAR} process bytes without inherent awareness of execution order semantics.

To tackle this, \textit{\sysname} utilizes a neural architecture that processes binaries by considering potential execution traces using a hybrid local-global attention mechanism. Specifically, it uses sliding window attention with special attention masks indicating execution reachability to capture local sequential information on the same trace, preventing focus on unrelated addresses. For global information, it adds a message-passing layer using an attention mechanism to pass information between jumps and calls, enabling the enforcement of long-range dependencies. This results in more contextually informed predictions aligned with CPU semantics.

The third challenge involves \textbf{enforcing consistency over probabilistic predictions.} Learning-based models inherently produce probabilistic outputs, assigning scores to potential instructions. However, a valid disassembly result must be a deterministic representation strictly adhering to the structural constraints. Simply thresholding instruction-level probabilities can lead to globally inconsistent results where individually plausible instructions collectively violate constraints.

\textit{\sysname} bridges this gap by incorporating a principled post-processing step. This framework leverages the previously characterized structural constraints (via the post-dominator tree) and employs a dynamic programming algorithm to prune inconsistencies from the model's probabilistic outputs. This reconciles neural scores with deterministic rules, ensuring a globally coherent and valid disassembly.

Our extensive evaluation on diverse datasets reveals a significant prevalence of structural constraint violations across disassembly tools. Neural disassemblers are particularly susceptible, exhibiting violations on most of the disassembly results. Surprisingly, even established rule-based disassemblers like IDA Pro and Ghidra are not immune, violating constraints on all binaries obfuscated with anti-disassembly techniques~\cite{Linn2003ObfuscationOE}. \textit{\sysname} significantly improves the consistency of disassemblers' outputs by completely eliminating all the structural constraint violations while maintaining high instruction-level accuracy.

Our primary contributions are as follows:

\noindent\textbf{Systematic Characterization of Structural Constraints.} We introduce a framework for characterizing structural constraints in disassembly and an efficient, PDT-based algorithm for detecting their violations without relying on the labels.

\noindent\textbf{Context-Aware Neural Disassembly Architecture.} We design a novel neural architecture that employs a hybrid attention mechanism, leading to more structurally informed disassembly results.

\noindent\textbf{Robust Enforcement of Deterministic Constraints.} We develop a principled post-processing technique to enforce the structural constraints. It applies to various disassemblers.

\noindent\textbf{Comprehensive Evaluation and Broader Impact.} Our extensive evaluation demonstrates that \textit{\sysname} maintains high accuracy while enforcing consistency. Furthermore, our error detection algorithm systematically locates many errors of widely-used  disassemblers and disassembly datasets.
 \section{Structural Constraints}
\label{sec: constraints}

\begin{figure*}[t]
\centering
\includegraphics[width=\linewidth]{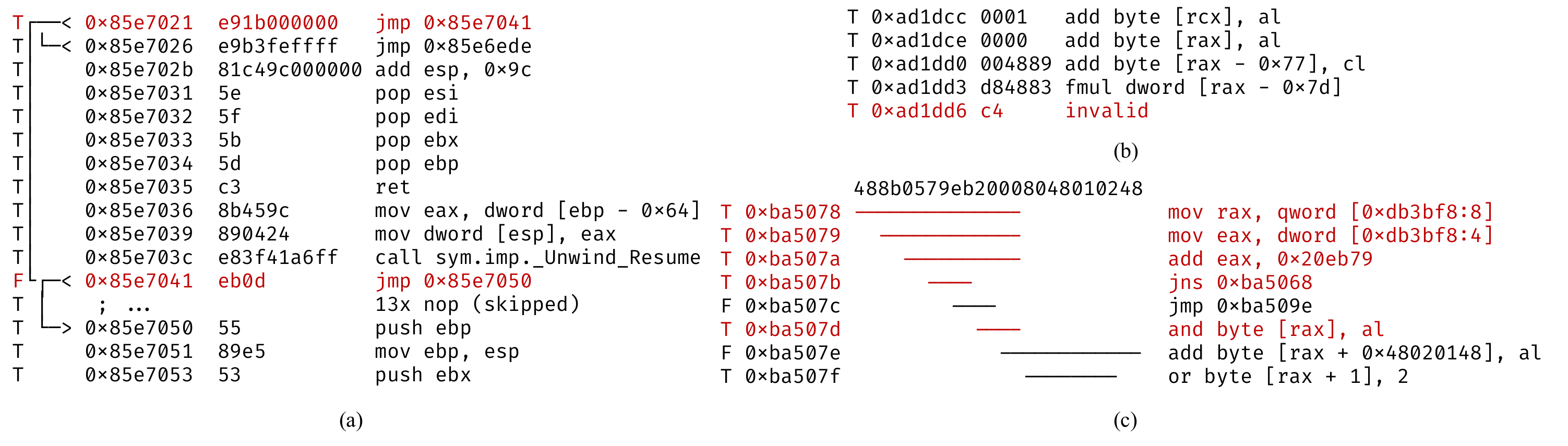}
\caption{Example of Constraint Violations discovered in the labels provided by x86-Sok dataset. (a) Missing Post-Dominator. (b) Dead-End Sequence. (c) Overlapping Instructions.}
\label{fig:case}
\end{figure*}
\subsection{Constraint Violations}

We identify three primary types of constraint violations. Each type is defined below, accompanied by an example identified within the labels provided by the x86-sok~\cite{Pang2020SoKAY} dataset. 

\noindent\textbf{Missing Post-Dominator (MPD).} This violation occurs when instruction $I_A$ identified as true code, but its post-dominator, instruction $I_B$, is missing from the disassembly. By definition of post-dominance, if $I_A$ is executed, then $I_B$ must also be executable. The incorrect labeling of $I_B$ thus constitutes a violation. Figure~\ref{fig:case}(a) illustrates an example from the binary \texttt{clang\_m32\_O0/ld.gold}. Here, the instruction at address \texttt{0x85e7021} is marked as true code. However, this unconditional jump instruction's target, \texttt{0x85e7041} (which is its post-dominator), is marked as non-code. This directly contradicts the post-dominance relationship implied by the jump instruction's semantics.

\noindent\textbf{Dead-End Sequence (DES).} This violation arises when an instruction $I_A$ is identified as true code, yet an undecodable byte sequence $U$ post-dominates $I_A$. Since every path from $I_A$ to a program exit must pass through $U$, and $U$ itself is not valid code, execution of $I_A$ would lead to an invalid state. Figure~\ref{fig:case}(b) presents an example from the binary \texttt{clang\_Of/libv8.so}. The instruction sequence starting from address \texttt{0xad1dcc} is labeled as code. This sequence, however, eventually leads to an undecodable byte sequence starting at \texttt{0xad1dd6}. This creates a dead-end, as execution cannot validly proceed through an undecodable address.

\noindent\textbf{Overlapping Instructions (OI).} This violation occurs when a disassembler proposes multiple distinct, valid instructions whose byte representations in memory overlap. The assumption that instructions do not overlap holds for most compiled code, with known exceptions typically limited to manually introduced lock prefixes via inline assembly or intentionally obfuscated code. A violation of this type usually indicates errors in the disassembly process. Figure~\ref{fig:case}(c) depicts an example from \texttt{clang\_O3/libv8.so}. The upper portion shows the hexadecimal representation of raw bytes, with annotations indicating the span of candidate instructions. In this instance, the addresses \texttt{0xba5078}, \texttt{0xba5079}, \texttt{0xba507a}, \texttt{0xba507b}, and \texttt{0xba507d} are all marked as true code. However, their byte sequences overlap. Manual investigation reveals that only the instruction at \texttt{0xba5078} is the actual true code.

It is important to emphasize that a real ground-truth disassembly should not exhibit these violations, as they fundamentally contradict the post-dominance relationships inherent in control flow semantics and the standard structure of executable code. However, correct disassembly can be complex and involve many subtle cases. Consequently, the labels provided by many disassembly datasets may contain errors and might not represent the real ground-truth. To avoid confusion, we consistently refer to such dataset entries as \texttt{labels} rather than \texttt{ground-truth}. Furthermore, the presence of these violations in the labels implies errors. This observation can be leveraged to systematically locate errors and improve the quality of disassembly datasets.

\subsection{Post-Dominator Tree}
\label{sec:PDT}

The post-dominance relation, crucial for the constraints discussed previously, is effectively characterized by the Post-Dominator Tree (PDT). In a PDT, nodes represent instructions from the Control Flow Graph (CFG), and the directed edges represent immediate post-dominance relationships.

A node $B$ is said to post-dominate node $A$ if every path from $A$ to an exit node in the CFG must pass through $B$. The immediate post-dominator of a node $A$ is its closest strict post-dominator. More formally, $B$ is the immediate post-dominator of $A$ if: (1) $B$ post-dominates $A$. (2) $B \neq A$. (3) There is no other node $P$ (where $P \neq A$ and $P \neq B$) such that $P$ post-dominates $A$ and $B$ post-dominates $P$.

\begin{figure}[!t]
    \centering
    \includegraphics[width=\linewidth]{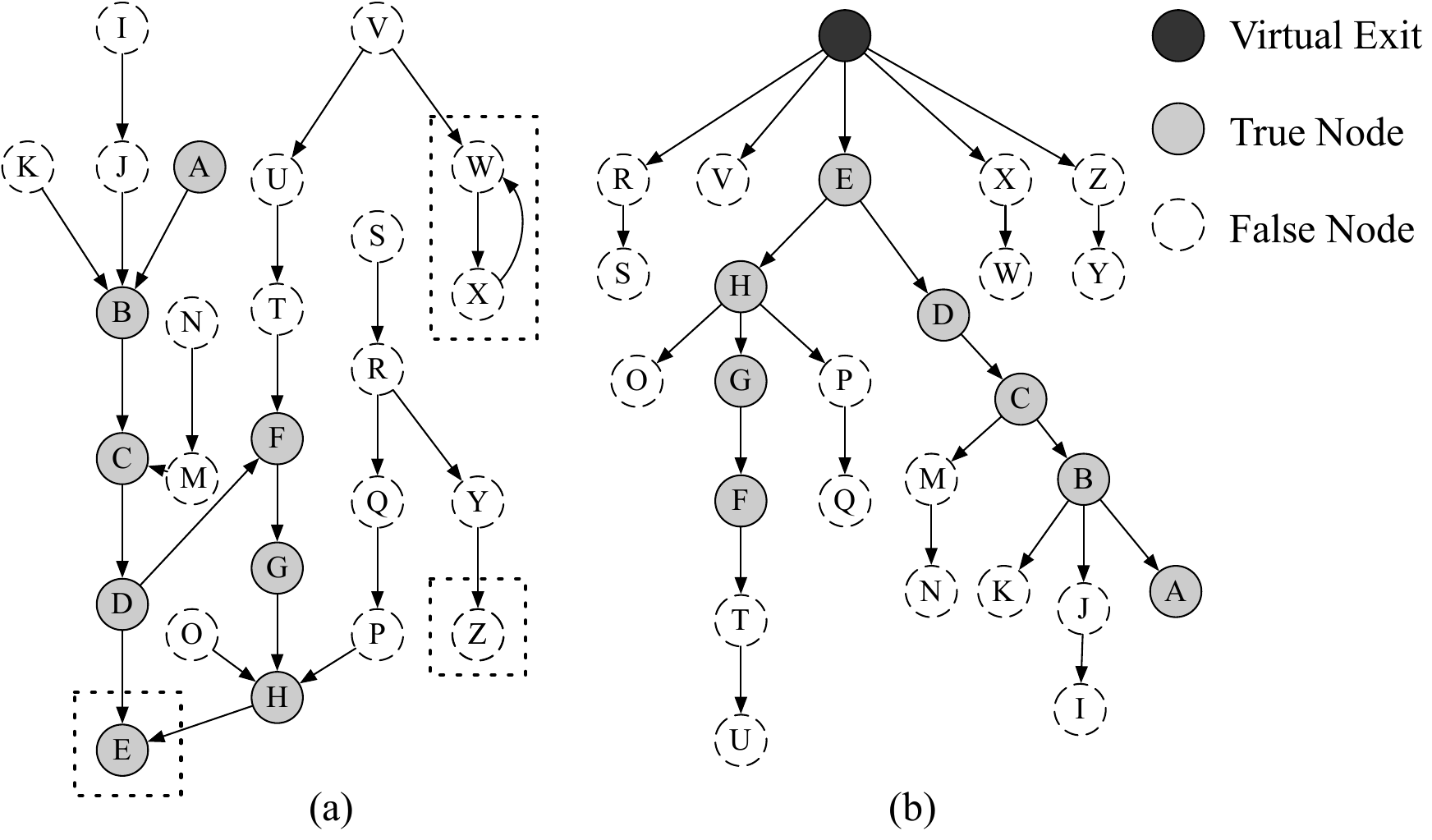}
    \caption{Example of a superset Control Flow Graph and its corresponding Post-Dominator Tree. (a) Superset Control Flow Graph. (b) Post-Dominator Tree.}
    \label{fig:pdt}
\end{figure}

Figure~\ref{fig:pdt} provides an illustrative example of a CFG and its corresponding PDT. In the depicted CFG, node $A$ is post-dominated by nodes $B, C, D,$ and $E$. Among these post-dominators, only node $B$ is the immediate post-dominator of $A$ according to the previous definition.

Constructing a PDT for a general CFG, particularly a superset CFG, which considers every address as potential instruction start and links all known control flow edges, presents unique challenges. Such CFGs may lack a single, common exit point and might not even form a connected graph. Therefore, a specialized construction process is necessary. Our PDT construction process is as follows.

\noindent\textbf{Handling Large-Scale Graphs with Weakly Connected Components (WCCs).}
Superset CFGs can be exceptionally large, potentially comprising millions of nodes, making direct processing infeasible due to memory and computational constraints. To manage this scale, we first segment the input CFG into its WCCs. A WCC is a maximal subgraph where a path exists between any two nodes if edge directions are disregarded. Since WCCs are disjoint, they can be processed independently and sequentially. This modular approach significantly decreases the memory footprint and computational burden, allowing our analysis to scale to very large binaries. To restrict the size of each WCC, we do not connect the edges from the \textit{call} instructions to their targets even if they can be determined statically. This approach restricts the sizes of the WCCs at function level, which is acceptable in practice. Otherwise, it is possible to find large WCCs consisting of almost all the functions in the binary. 

\noindent\textbf{Establishing a Unified Exit Point for each WCC.}
After decomposing the CFG into WCCs, each WCC is, by definition, connected (when viewed as undirected). However, a WCC may still possess multiple natural exit points, such as multiple \texttt{ret} instructions or program termination syscalls. To ensure that the post-dominance relation within each WCC forms a tree structure with a single root, we introduce a virtual exit node specific to that WCC. This virtual exit node becomes the sole successor for all nodes within the WCC that originally had an out-degree of zero.

\noindent\textbf{Managing Cycles with Strongly Connected Components~(SCCs).}
A SCC is a maximal subgraph where there is a directed path from any node to any other node within that SCC. Infinite loops or complex cyclic structures within a WCC present a challenge for post-dominance analysis, as nodes within such cycles might not have a clear path to an external exit if all their successors are also within the cycle. We resolve this by identifying the SCCs of the WCC. The WCC can then be conceptually regarded as a condensation graph, where each SCC is collapsed into a single node; this condensation graph is inherently acyclic.

Instead of connecting all original zero out-degree nodes of the WCC directly to its virtual exit, we refine this for terminal SCCs. A terminal SCC is one that has no outgoing edges to nodes outside of itself within the WCC. It has an out-degree of zero in the condensation graph of the WCC. For each such terminal SCC, we connect designated representative nodes from within that SCC to the WCC's virtual exit node. For our purposes, these representative nodes are the control-flow transfer instructions within the respective SCCs that create the loops, which are the jumps. This ensures that nodes within terminal loops are properly anchored in the PDT. For instance, in Figure~\ref{fig:pdt}, the rectangle enclosing $E$ and $Z$ are single node terminal SCC, which are directly linked to the virtual exit. Node $W$ and $X$ form a terminal SCC representing an infinite loop, and the jump instruction $X$ is chosen as the representative node to link to the virtual exit.

\noindent\textbf{PDT Generation.}
With each WCC preprocessed as described, augmented with its own virtual exit node and with its terminal SCCs appropriately linked to this exit, we then compute the immediate post-dominators for all nodes within that WCC. This is achieved using a standard algorithm, such as the Lengauer-Tarjan algorithm~\cite{Lengauer1979AFA} on the reversed graph, since it is originally used to calculate immediate dominator instead of immediate post-dominator. The collection of these individual trees (one for each WCC, rooted at its virtual exit) forms the overall PDT for the input CFG. 

It is important to note that our PDT construction does not require a fully reconstructed or perfectly complete CFG; it can operate effectively even with partial control-flow information, such as when the targets of some indirect jumps or calls remain unresolved.

\subsection{Detection of Structural Violations}
\label{sec:pdt_violation_detection}

The PDT derived from a superset CFG provides a powerful structure for efficiently validating disassembly consistency. A correctly disassembled program, when represented as a PDT, exhibits specific structural properties. Primarily, all nodes representing actual instructions (true nodes) must form a connected subtree rooted at the virtual program exit. This implies the following two fundamental conditions.

\noindent\textbf{Path Integrity.} For any true node, the entire path from that node to the PDT root must consist exclusively of true nodes. Consequently, no true node should be post-dominated by a false node (decodable but classified as false) or an invalid node (a byte sequence that is not decodable).

\noindent\textbf{Non-Overlapping Assumption.} Adhering to the standard assumption that instructions do not overlap, a node in the PDT cannot post-dominate multiple true child nodes if those children represent non-control-flow (NCF) instructions. If it did, it would imply that multiple distinct NCF instructions fall-through to the same subsequent instruction, which is only possible if they overlap in memory.

\begin{figure}[!htbp]
    \centering
    \includegraphics[width=0.9\linewidth]{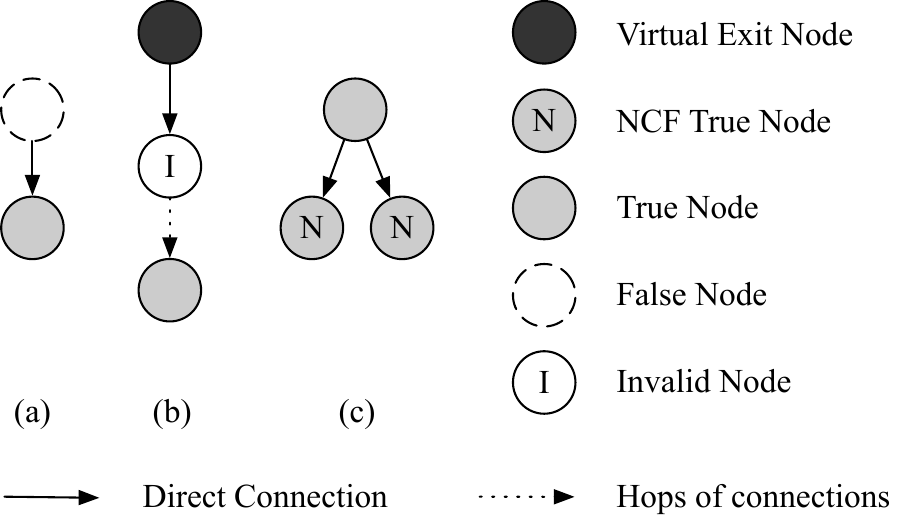}
    \caption{Patterns of violations. (a) Missing Post-Dominator. (b) Dead-End Sequence. (c) Overlapping Instructions.}
    \label{fig:tree_property}
\end{figure}

Violations of these properties appear as identifiable patterns within the PDT. Figure~\ref{fig:tree_property} illustrates these patterns. Our detection algorithm, presented in Algorithm~\ref{alg:detection}, systematically identifies these patterns through a single traversal of the PDT and returns three error sets.

\begin{algorithm}[t]
    \caption{Structural Violations Detection}
    \label{alg:detection}
    \footnotesize
    \begin{algorithmic}[1] 
        \Procedure{Error Detection}{$PDT$, $r$} \Comment{$r$ is the PDT root}
            \State Initialize error sets $E_{1} \gets \emptyset$, $E_{2} \gets \emptyset$, $E_{3} \gets \emptyset$ and $visitedSet \gets \emptyset$
            \State Start BFS from root $r$, add $r$ to $queue$ and $visitedSet$

            \Function{DetectDeadEnd}{node}
                 \State Find all descendants of $node$ marked true via DFS
                 \State Add them to $E_{2}$ \Comment{Dead-End Sequence}
            \EndFunction
            
            \While{BFS $queue$ not empty}
                \State $node \gets$ next node from $queue$
                \If{$node$ is True or is root}
                    \For{each child $c$ of $node$}
                        \State Add $c$ to $visitedSet$
                        \If{$c$ is non-control-flow}
                            \If{$node$ is root}
                                \State DetectDeadEnd($c$)
                            \ElsIf{$c$ is True and $node$ visited True NCF child}
                                \State Add $c$ to $E_{3}$ \Comment{Overlapping Instructions}
                            \EndIf
                        \Else
                            \If{$c$ is True and $c \notin visitedSet$}
                                \State Add $c$ to BFS $queue$
                            \EndIf
                        \EndIf
                    \EndFor
                \EndIf
            \EndWhile

            \For{each node $v$ in $PDT$}
                 \If{$v \notin visitedSet$}
                    \State Add $v$ to $E_{1}$ \Comment{Missing Post-Dominator}
                 \EndIf
            \EndFor
            
            \State \Return $(E_{1}, E_{2}, E_{3})$
        \EndProcedure
    \end{algorithmic}
\end{algorithm}

\noindent\textbf{Missing Post-Dominator ($E_{1}$).} This violation occurs when a true node is post-dominated by a false node, as shown in Figure~\ref{fig:tree_property}(a). The true node's path to the PDT root contains a false node, violating the \textit{path integrity} property. The algorithm performs a BFS (lines 3, 8-26) to populate a \textit{visitedSet} with all true nodes reachable from the root $r$ through paths of true nodes. After the BFS, any true node $v$ in the PDT that is not visited is added to $E_1$ (lines 27-31).

\noindent\textbf{Dead-End Sequence ($E_{2}$).} This occurs if a true instruction is post-dominated by an invalid node, as shown in Figure~\ref{fig:tree_property}(b). The helper function (lines 4-7) is called for each NCF child $c$ of the root node (line 15). All such NCF nodes are invalid nodes, since otherwise they should be post-dominated by their subsequent instruction and cannot be immediate post-dominated by the virtual exit. This function performs a DFS from $c$ to collect all its true descendants to $E_2$. 

\noindent\textbf{Overlapping Instructions ($E_{3}$).} This violation occurs if a node has multiple true NCF children, as shown in Figure~\ref{fig:tree_property}(c). This implies that the instructions overlap with each other. During the BFS, if a node (that is not the root) has a true NCF child $c$, and the node has already visited a True NCF child, then $c$ is added to $E_3$ (lines 16-18).

By identifying these characteristic patterns, our approach detects violations of structural constraints from the PDT.

 \section{Tady}

\begin{figure}[!htbp]
    \centering
    \includegraphics[width=0.95\linewidth]{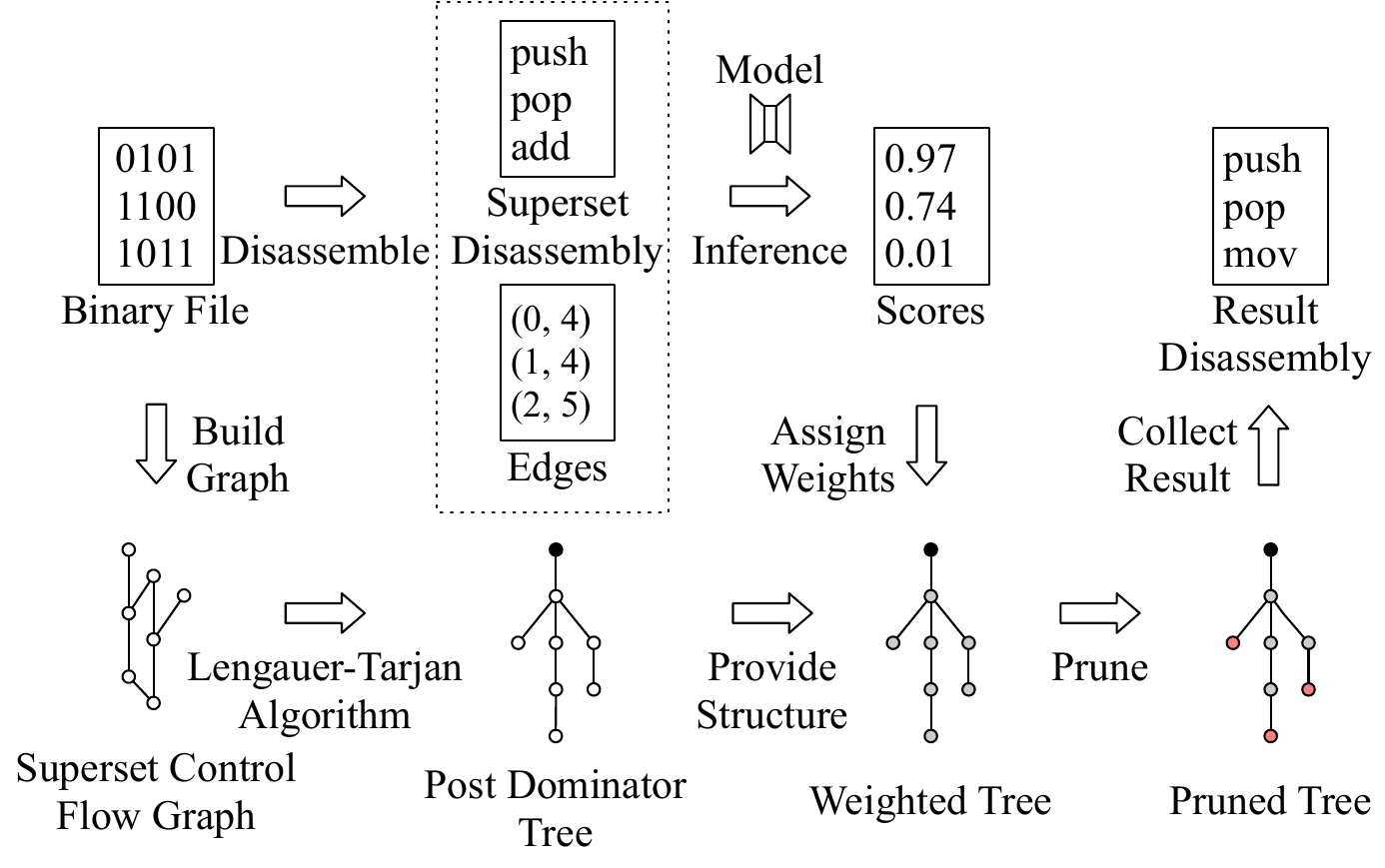}
    \caption{Overview of Tady.}
    \label{fig:overview}
\end{figure}

Figure \ref{fig:overview} shows our disassembler's workflow. We first conduct a superset disassembly over the executable section to extract instructions and their address connections. They are then fed into our model to assign scores to each address. Simultaneously, we construct a superset CFG connecting all instructions. The nodes are all instructions from the superset disassembly and the edges are the control flow edges, including fall-through, jump, conditional-jump and call. 

We convert this CFG into a PDT with the steps described in Section~\ref{sec:PDT}. Each node is then assigned a weight based on the model's prediction, representing its likelihood of being a true instruction. After assigning the weights to PDT nodes, we apply our post-processing algorithm to regularize the solution. The inconsistent nodes are pruned and missing internal nodes are recalled. The nodes that remain on the tree after pruning form the final disassembly result.

In the following subsections, we first explain how our trace-aware model design incorporate necessary context to prevent constraint violations. Then, we introduce the algorithm for post-processing the output of model to enforce the result to satisfy the constraints.

\subsection{Model Design}

Satisfying the diverse constraints inherent in accurate instruction recognition, as detailed in Section~\ref{sec: constraints}, necessitates a model capable of processing both local and global contextual information. While non-overlapping assumptions primarily require local context regarding immediately adjacent instructions, constraints based on post-dominance demand a broader understanding of instruction interdependencies, often spanning considerable distances within the code.

Previous approaches exhibit limitations in addressing these multifaceted requirements. DeepDi~\cite{Yu2022DeepDiLA}, for instance, utilizes a Graph Neural Network (GNN) to propagate information along the execution trace. However, the GNN's receptive field, constrained by its layer-wise message passing, struggles to capture long-range dependencies critical for post-dominance constraints, where related instructions can be many hops apart. Conversely, XDA~\cite{Pei2020XDAAR} processes raw byte sequences without explicitly leveraging control flow semantics. This makes it difficult for the model to effectively reason about control-flow-dependent constraints.

\begin{figure}[!bp]
    \centering
    \vspace{-20pt}
    \includegraphics[width=0.6\linewidth]{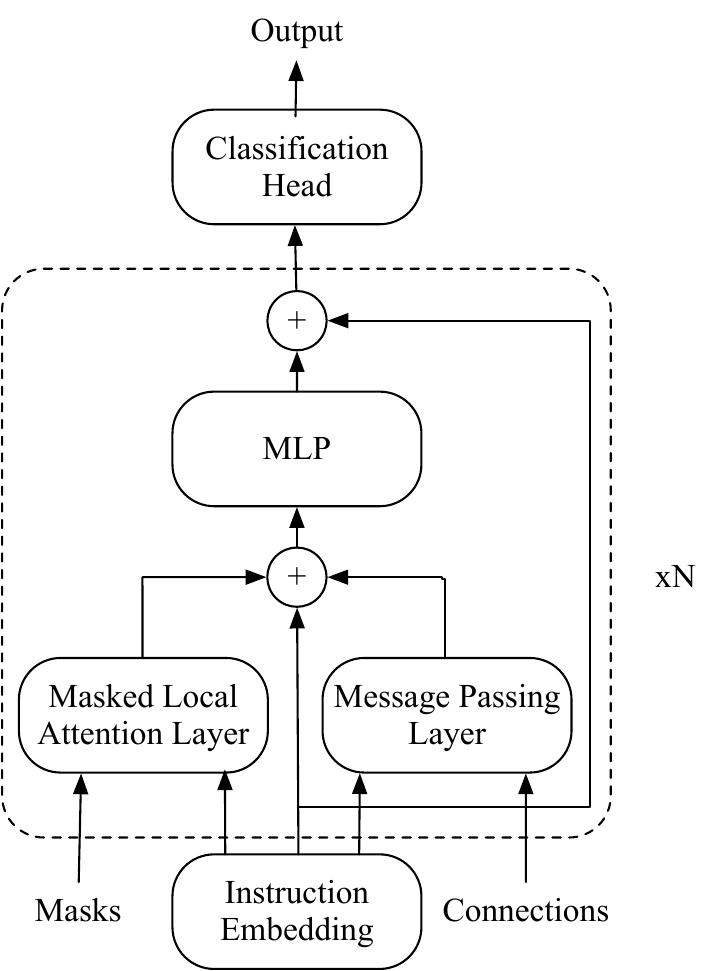}
    \caption{Overview of the Model Design.}
    \label{fig:model_design}
\end{figure}

Our model introduces a novel architecture specifically designed to overcome these challenges. It integrates two key mechanisms.
First, to capture fine-grained local context, we employ a masked local attention mechanism. This involves a transformer architecture equipped with a sliding window attention mechanism, notably augmented by an attention mask. This mask filters out instructions that, despite being within the sliding window, are not relevant to the task, thereby ensuring that local attention focuses only on semantically relevant preceding and succeeding instructions.

Second, to address the limitations of purely local attention and capture distant dependencies, we incorporate a global message passing layer. This layer facilitates information exchange between instruction blocks that may be spatially distant in the static code but are connected through control flow. This allows the model to effectively learn relationships critical for satisfying constraints like post-dominance, which often involve instructions far apart in the disassembly but proximate on potential execution paths.

As illustrated in Figure~\ref{fig:model_design}, our model processes input as follows:
First, an RNN layer generates initial instruction-level embeddings from the input instruction sequence. These embeddings then pass through a series of transformer blocks. Each transformer block uniquely combines the masked local attention mechanism (for precise local context) and the global message-passing layer (for capturing long-range, control-flow-aware dependencies). Finally, a classification head processes the refined hidden states from the transformer blocks, outputting a score for each potential instruction start address, indicating its likelihood of being a true instruction.

\subsubsection{Instruction Embedding}
\label{sec:instr_emb}
Instruction-level embeddings are generated with efficiency as a primary consideration, given that disassembly is a performance-critical task. We have empirically found that using raw bytes directly, rather than more complex tokenization over printable assembly, provides lightweight yet sufficiently effective features for this purpose. 

Initially, the model is provided with the raw byte sequence of the input code and the decoded length of each instruction. Within the model, the specific byte sequence for each individual instruction is recovered using this information. 

These bytes are then converted into numerical IDs. To preserve positional information within an instruction, the ID for a byte at offset i within that instruction is augmented by adding $256i$. Given that the maximum instruction length for x86-64 is 15 bytes, this scheme results in $256\times15=3840$ unique possible IDs. These numerical IDs are then passed through an embedding layer to create dense vector representations for each byte. This sequence of byte embeddings for each instruction is then processed by an RNN layer to compute the final instruction-level embedding.

\subsubsection{Masked Sliding Window Attention}
\begin{figure}[!bp]
\centering
\vspace{-10pt}
\includegraphics[width=0.9\linewidth]{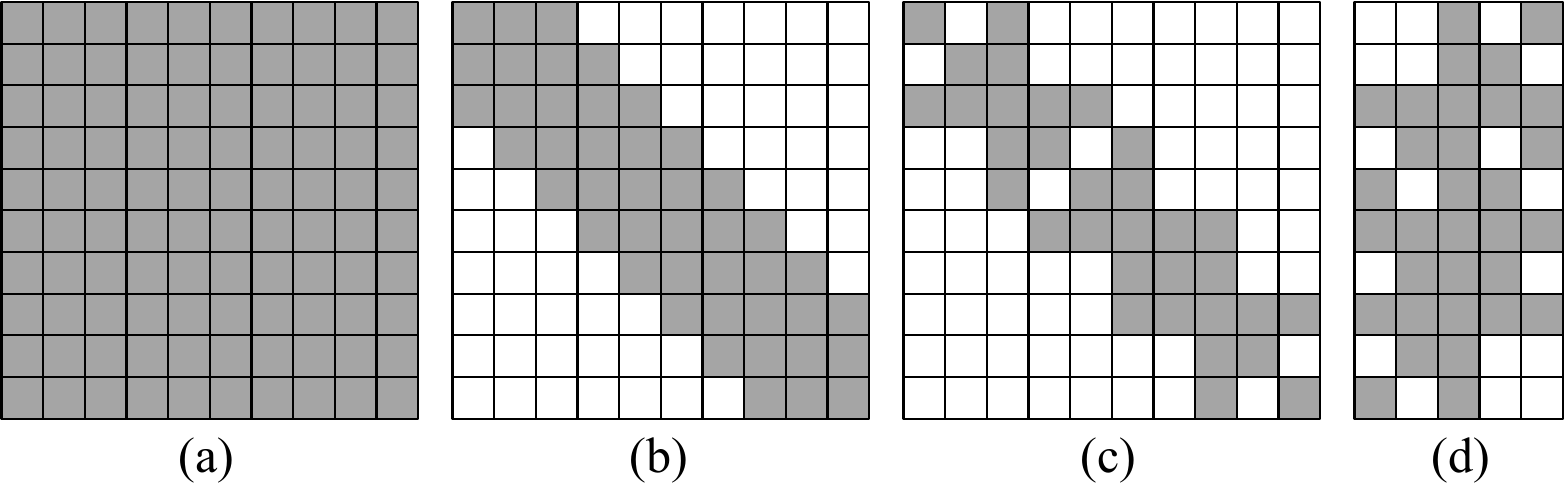}
\vspace{-10pt}
\caption{Sliding Window Attention with an Example Reachability Mask. (a) Full Attention. (b) Sliding Window Attention. (c) Masked Sliding Window Attention. (d) Flattened Mask.}
\label{fig:attention_mask}
\end{figure}

We introduce Masked Sliding Window Attention (MSWA) as a mechanism for capturing local context information. This approach constrains each instruction to attend only to a relevant subset of instructions within a sliding window, defined by a specific attention mask, as shown in Figure~\ref{fig:attention_mask}(c). 

Furthermore, this masking strategy can be applied heterogeneously within a single multi-head attention layer: different attention heads can utilize distinct masks, allowing them to specialize in capturing different types of relationships simultaneously. For instance, an execution-order context is captured using a reachability mask, which limits attention to instructions within the same execution trace. Another application is an overlapping relation context, where the mask allows an instruction to attend to all other instructions it overlaps with. 

Within each sliding window, as shown in Figure~\ref{fig:attention_mask}(b), an instruction's attention is restricted based on the applied mask. For computational efficiency, the sliding window pattern is typically flattened, as depicted in Figure~\ref{fig:attention_mask}(d), requiring much less computation comparing to the full attention as shown in Figure~\ref{fig:attention_mask}(a). The masked attention computation for a window of instructions is defined as:

$$\text{Attention}(Q,K_{s},V_{s}) = \text{softmax}\left(\frac{Q(K_{s})^{T}}{\sqrt{d_{k}}} + M_t \right)V_{s}$$

where Q, K, and V are the query, key, and value matrices respectively. The subscript $s$ indicates that keys and values are selected within the sliding window. The mask $M_t$ applies large negative values to disallowed attention pairs, effectively ensuring near-zero attention weights between instructions that do not satisfy the specific contextual relationship $t$.

Two variants of attention mask are used in our model. The \textbf{reachability mask} is derived from the connections information parsed from the instructions' semantics. To determine reachability between instructions, we follow each instruction's execution in parallel until it reaches maximum steps and collect reachable instructions along the way. The collection stops when encountering conditional jumps for simplicity. The \textbf{overlapping mask} is determined by identifying all instructions within the window that overlap with the current instruction, which can be calculated based on the given input length information as mentioned in Section~\ref{sec:instr_emb}.

Our MSWA framework offers significant advantages by enabling focused attention based on defined local contexts. It allows direct attention between all instructions on the same execution path within a window, rather than requiring multiple message-passing iterations as in GNNs. For instance, in a sequence $A\to B\to C$, the reachability mask allows C to directly attend to both A and B in a single layer. 

The applied mask provides attention that is more focused compared to naive sequence models by filtering out instructions irrelevant to the specific local context being modeled, ensuring attention is concentrated on valid or pertinent relationships. The ability to assign different masks to different attention heads further enhances this focused attention, allowing the model to simultaneously learn diverse contextual features within a single layer.

\subsubsection{Global Message Passing}
\begin{figure}[!t]
    \centering
\includegraphics[width=\linewidth]{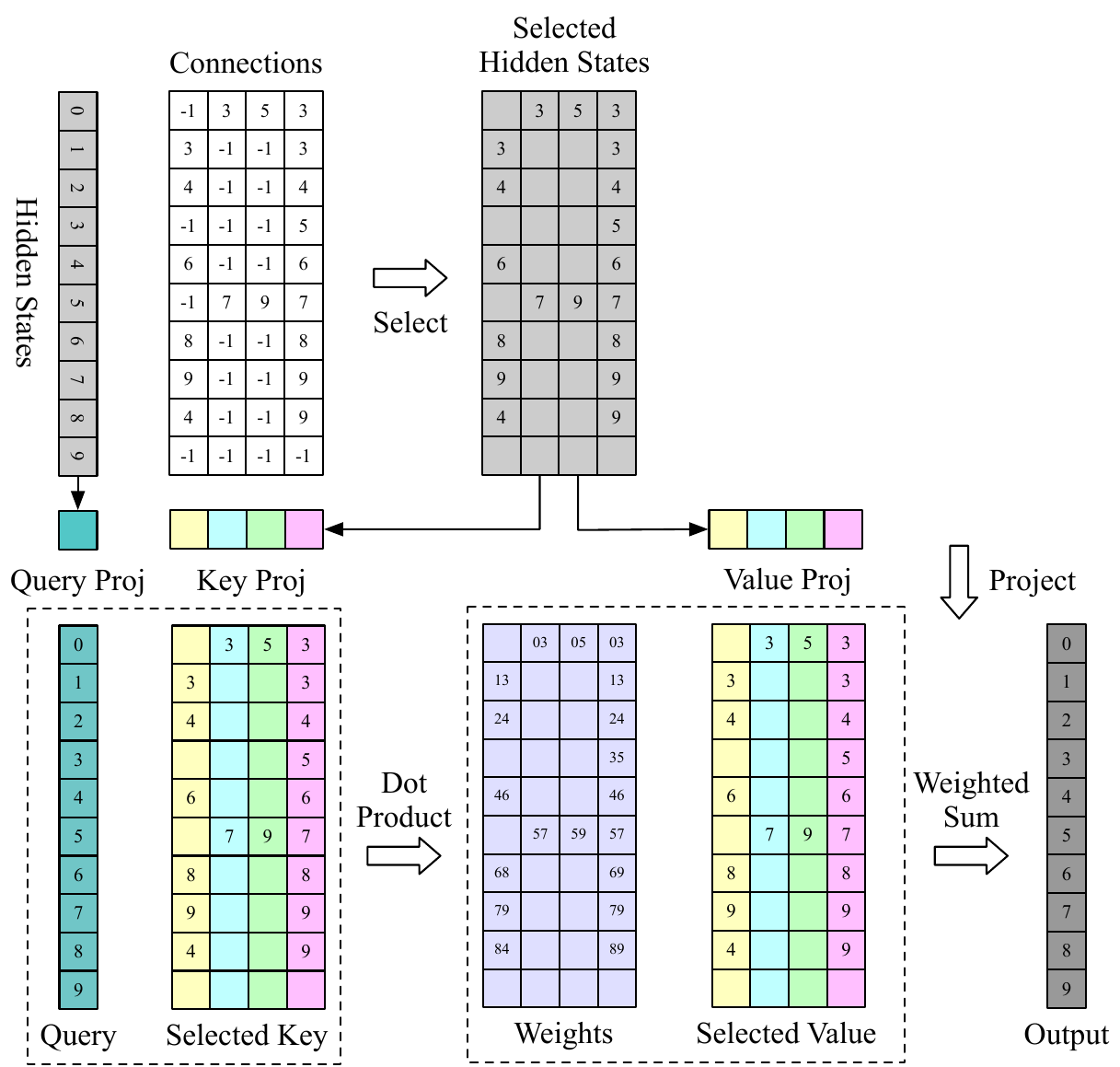}
    \vspace{-10pt}
    \caption{Global Message Passing With Selective Attention.}
    \label{fig:selective_attention}
\end{figure}

To effectively capture the global structure of a program, we introduce a \textbf{global message passing mechanism}. This mechanism leverages selective attention across various types of instruction relationships. For each instruction node $i$, we categorize its connections as follows:

\noindent\textbf{Must Transfer ($M_{i}$)}: The unique successor instruction that is guaranteed to execute immediately after instruction $i$.

\noindent\textbf{May Transfer ($T_{i}$)}: Both the potential jump target and the fall-through instruction of a conditional branch instruction.

\noindent\textbf{Next ($N_{i}$)}: The instruction that sequentially follows instruction $i$ in the program's memory layout.

The total number of these connections for any given node $i$ is $C = |M_{i}| + |T_{i}| + |N_{i}| = 4$. These connections define an adjacency structure, represented as a tensor of shape $(B, L, C)$, where $B$ is the batch size and $L$ is the sequence length.

As illustrated in Figure~\ref{fig:selective_attention}, for each instruction $i$, represented by its hidden state $h_i$, we compute attention scores over its neighboring instructions using type-specific projections:
\[
    Q_i = W_q h_i
\]
\[
    K_{j,t} = W_k^t h_j \quad \text{for } j \in \{M_i \cup T_i \cup N_i\}
\]
\[
    V_{j,t} = W_v^t h_j \quad \text{for } j \in \{M_i \cup T_i \cup N_i\}
\]
Here, $t$ denotes the specific connection type ($M$, $T$, or $N$). $W_q$, $W_k^t$, and $W_v^t$ are learnable projection matrices.

The attention weights $\alpha_{i,j,t}$ between instruction $i$ and a neighbor $j$ of type $t$ are computed using a softmax function:
\[
    \alpha_{i,j,t} = \text{softmax}\left(\frac{Q_i K_{j,t}^T}{\sqrt{d_k}}\right)
\]
where $d_k$ is the dimension of the key vectors, used for scaling.

The updated hidden state $h'_i$ for instruction $i$ is calculated as a weighted sum of the value vectors from its neighbors:
\[
    h'_i = \sum_{t \in \{M,T,N\}} \sum_{j \in \text{type } t \text{ neighbors of } i} \alpha_{i,j,t} V_{j,t}
\]

Some instructions lack certain connection types. For example, an unconditional jump has no \textit{May Transfer} edges. We employ masked attention for them, setting the corresponding attention weights $\alpha_{i,j,t}$ to zero for non-existent connections, effectively preventing information flow through those paths.

This selective attention mechanism facilitates efficient information propagation across the program's control flow graph. It is effective for relaying information between distant instructions connected via jumps or calls. This contributes to maintaining global consistency in the result.

\subsection{Pruning Algorithm}

While our model's rich feature set helps satisfy constraints, it cannot guarantee full compliance with the hard constraints of valid disassembly. Post-processing is therefore essential to ensure validity of the final output. Since these constraints are fundamentally tied to post-dominance relation, we develop a post-processing algorithm based on PDT. As illustrated in Figure \ref{fig:pdt}, our goal is to find a subset of nodes in the PDT that satisfies the properties discussed in Section \ref{sec:pdt_violation_detection}, while maximizing the sum of their confidence scores. This can be viewed as finding the most probable self-consistent disassembly solution. We formulate this as a dynamic programming problem where we seek to find a maximum-weighted subtree rooted at the original PDT root. The problem naturally decomposes into sub-problems of finding optimal subtrees rooted at each child node. By solving these sub-problems recursively and combining their solutions, we obtain the globally optimal pruned tree. Our pruning algorithm consists of two phases.

\noindent\textbf{Weight Propagation.} The Weight Propagation phase, detailed in Algorithm~\ref{alg:weight_propagation}, aggregates instruction-level confidence scores upward through the tree to compute subtree weights. This prevents naive greedy pruning that might discard entire subtrees due to a single negative-weighted node, even when the subtree contains many high-confidence instructions.

\begin{algorithm}[t]
    \caption{Weight Propagation in Post-Dominator Tree}
    \label{alg:weight_propagation} \footnotesize
    \begin{algorithmic}
        [1] \Procedure{PropagateWeights}{$tree, root$} \Function{PostOrderVisit}{$node$}
        \State $childW \gets 0$ \For{$child \in \text{Children}(node)$} \State
        \Call{PostOrderVisit}{$child$} \If{$weight[child] > 0$} \State $childW \gets
        childW + weight[child]$ \EndIf \EndFor \State $weight[node] \gets \max(0,
        weight[node] + childW)$ \EndFunction \State \Call{PostOrderVisit}{$root$}
        \EndProcedure
    \end{algorithmic}
\end{algorithm}

As shown in Algorithm~\ref{alg:weight_propagation}, we perform a post-order depth-first traversal where each node's weight is updated only after processing all its children. The key insight is that we selectively aggregate only positive weights from children (lines 6-8), ensuring the pruning decisions maximize the overall confidence of the retained instructions. The final weight of each node combines its own score with the aggregated child weights, clamped to be non-negative (line 10).

\begin{algorithm}[ht]
    \caption{Pruning Post-Dominator Tree}
    \label{alg:pruning} \footnotesize
    \begin{algorithmic}
        [1] \Procedure{PruneTree}{$tree, root$} \State $prunedTree \gets \emptyset$
        \State $Q \gets \text{new Queue}()$ \State $Q.\text{push}(root)$ \While{$Q \text{ is not empty}$}
        \State $node \gets Q.\text{pop}()$ \If{$weight[node] > 0 \text{ or }node = root$}
        \State $\text{AddNode}(prunedTree, node)$ \State $maxFT \gets (0, \text{null}
        )$ \For{$child \in \text{Children}(node)$} \If{$\text{not IsControlFlow}(child)$}
        \If{$weight[child] > maxFT.w$} \State
        $maxFT \gets (weight[child], child)$ \EndIf \ElsIf{$weight[child] > 0$} \State
        $\text{AddNode}(prunedTree, child)$ \State $\text{AddEdge}(prunedTree, no
        de, child)$ \State $Q.\text{push}(child)$ \EndIf \EndFor \If{$maxFT.n \neq \text{null}$}
        \State $\text{AddNode}(prunedTree, maxFT.n)$ \State
        $\text{AddEdge}(prunedTree, node, maxFT.n)$ \State
        $Q.\text{push}(maxFT.n)$ \EndIf \EndIf \EndWhile \State \Return
        $prunedTree$ \EndProcedure
    \end{algorithmic}
\end{algorithm}

\noindent\textbf{Result Collection.} The Result Collection phase, implemented in Algorithm \ref{alg:pruning}, constructs the final pruned tree through a breadth-first traversal. Starting from the root, we build the pruned tree by selectively including nodes based on their propagated weights and types.

The algorithm maintains two key invariants: (1) Control-flow instructions are included only if they have positive weights (lines 15-19). (2) For non-control-flow children of each node, only the highest-weighted child is retained (lines 11-15, 21-25). This selective inclusion, combined with the breadth-first traversal order, ensures the resulting tree maintains all constraints while pruning away low-confidence or inconsistent instructions. The final pruned tree, returned on line 28, represents a valid disassembly solution that maximizes confidence scores while satisfying all structural constraints.

 \section{Evaluation}
In this section, we evaluate four key research questions:

\noindent\textbf{RQ1 (Prevalence of Inconsistency).} What is the frequency and severity of structural constraint violations in disassemblers' results and datasets' labels?

\noindent\textbf{RQ2 (Disassembly Performance).} How well does our neural disassembler perform compared to other approaches?

\noindent\textbf{RQ3 (Efficiency).} How efficient are our model, the error detection and the pruning algorithms?

\noindent\textbf{RQ4 (Ablation Study).} How effectively do the aspects of our model design contribute to the overall performance?

\subsection{Implementation and Setup}

\noindent\textbf{Implementation.} For instruction decoding and the extraction of associated metadata, including byte length, control flow characteristics, and successor instructions, we utilize the MCDisassembler from the LLVM compiler infrastructure~\cite{lattnerLLVMCompilationFramework2004}. To optimize for processing efficiency, we directly use the internal representation of instructions rather than their printable, objdump-style disassembly output. The error detection and pruning algorithms are implemented in C++ and leverage the Boost Graph Library. To facilitate integration with Python-based workflows, these C++ implementations are equipped with Python bindings that expose results as NumPy arrays. Neural network models are implemented and trained using Flax~\cite{flax2020github}, which operates on top of JAX~\cite{jax2018github}. For operational deployment, trained models are exported to the TensorFlow~\cite{tensorflow2015-whitepaper} SavedModel format. Serving is then managed by TensorFlow-Serving~\cite{Olston2017TensorFlowServingFH}, with a Python client responsible for data preprocessing and gRPC communication with the serving endpoint. All experiments were conducted on a PC equipped with an Intel Core i9-12900K CPU, a single NVIDIA RTX A6000 Ada generation GPU, 64GB of RAM, and a 1TB NVMe SSD.

\noindent\textbf{Deep Learning Model Settings.} For the optimization of our deep learning model, we utilize the AdamW optimizer~\cite{Loshchilov2017DecoupledWD} with a learning rate of $1 \times 10^{-3}$. To address the issue of class imbalance in the training data, we employ Focal loss~\cite{Lin2017FocalLF}, a variant of Binary Cross-Entropy loss~\cite{Goodfellow-et-al-2016} designed to prioritize difficult examples. The specific hyper-parameters for the Focal loss are set to $\alpha = 0.8$ and $\gamma = 4.0$. Our model architecture consists of two transformer layers incorporating our custom attention mechanism. The hidden size of the transformer layers is 16, and the intermediate size is 32. We use a sliding window approach with a window size of 64 on both sides. We use four attention heads for our MSWA, three of them use reachability mask and the other one uses overlapping mask. The input sequences are chunked into pieces of length 8192; chunks shorter than this maximum are padded with zeros to ensure consistent input dimensions.

\noindent\textbf{Datasets.}
Our experiments utilize several public benchmarks. For training, we primarily use Pangine~\cite{Li2020OnTG}, a dataset of 879 binaries compiled with clang~(3.8, 6.0), gcc~(5.4, 7.0), icc~(19.1.1.219), and msvc-cl~(19.26.28806) across various optimization levels (O0-O3, Os, Ofast) for x86 and x86-64 architectures. Its labels are derived from intermediate compilation results. The model is trained on this dataset for one epoch with a 9:1 train-validation split. The trained model is evaluated on Assemblage~\cite{Liu2024AssemblageAB} (large-scale open-source Windows programs from GitHub, labels from PDB files) and x86-sok~\cite{Pang2020SoKAY} (Linux binaries, including complex ones like openssl and mysqld, with labels from modified clang 6.0.0 and gcc 8.1). We excluded bap-corpora~\cite{baoBYTEWEIGHTLearningRecognize2014} as its components are covered by x86-sok. Additionally, we created rw, a real-world dataset mirroring DeepDi's~\cite{Yu2022DeepDiLA} protocol, using the latest versions of their selected projects and x86-sok's modified compilers for labels and binaries. Acknowledging limitations in label generation (e.g., incomplete coverage from external linking, ambiguity of padding bytes), we follow ddisasm-WIS~\cite{flores-montoyaDisassemblyWeightedInterval2025a} by marking addresses without labels as \texttt{ignore}, excluding them from training and evaluation. All experiments focus on the \texttt{.text} section.

To assess robustness against obfuscation, we use OLLVM-14.0~\cite{junodObfuscatorLLVMSoftwareProtection2015} (LLVM IR level), Tigress~\cite{banescuCodeObfuscationSymbolic2016} (source code level), and binobf~\cite{Linn2003ObfuscationOE} (binary-level anti-disassembly). For binobf, we use the 11 SPEC CPU 2000 binaries from the obf-benchmark~\cite{Linn2003ObfuscationOE} dataset. For OLLVM and Tigress, we use the quarks dataset~\cite{quarkslabDiffing_obfuscation_dataset2025}, which includes variously configured obfuscated binaries. There is a obfuscation ratio controlling how many functions are obfuscated within the binary, and we choose the subset of the dataset where this ratio is 100\%, indicating all the functions are obfuscated. Details of the quarks dataset's composition can be found in its documents.

Our final evaluation set includes 81,875 binaries from Assemblage, 3,997 from x86-sok, 879 from Pangine, 526 from rw, 1,298 from quarks, and 11 from obf-benchmark. To manage evaluation time, datasets with over 1,000 binaries are randomly sampled (1,000 binaries, seed 0); smaller datasets are used entirely. For rule-based disassemblers, a 60-second timeout per binary is enforced; binaries causing failures or timeouts are excluded from that disassembler's results.

\begin{table*}[!ht]
\vspace{-5pt}
\centering \caption{Comparison of Disassembly Tools Across Datasets and Error Metrics (F: File Error Rate, B: Errors per 1MB).}
\label{tab:error} \sisetup{
  round-mode=places, round-precision=4, table-align-exponent = false, }
\setlength{\tabcolsep}{3pt}
\begin{tabular}{@{}cc *{7}{S[table-format=1.2, round-precision=2]S[table-format=5.1, round-precision=1]}@{}}\toprule
\multirow{2}{*}{Dataset} & \multirow{2}{*}{Err} & \multicolumn{2}{c}{Labels} & \multicolumn{2}{c}{IDA} & \multicolumn{2}{c}{Ghidra} & \multicolumn{2}{c}{Ddisasm} & \multicolumn{2}{c}{DeepDi} & \multicolumn{2}{c}{XDA} & \multicolumn{2}{c}{Tady} \\
\cmidrule(lr){3-4} \cmidrule(lr){5-6} \cmidrule(lr){7-8} \cmidrule(lr){9-10} \cmidrule(lr){11-12} \cmidrule(lr){13-14} \cmidrule(lr){15-16}
 &  & {F} & {B} & {F} & {B} & {F} & {B} & {F} & {B} & {F} & {B} & {F} & {B} & {F} & {B} \\
\midrule
\multirow{5}{*}{Pangine} & M & 0.028 & 0.6 & 0.014 & 0.0 & 0.007 & 0.1 & 0.000 & 0.0 & 0.937 & 60.9 & 1.000 & 7455.8 & 0.879 & 22.7 \\
 & D & 0.000 & 0.0 & 0.009 & 0.0 & 0.009 & 0.1 & 0.001 & 0.0 & 0.596 & 11.2 & 0.993 & 483.3 & 0.000 & 0.0 \\
 & O & 0.003 & 0.0 & 0.000 & 0.0 & 0.000 & 0.0 & 0.000 & 0.0 & 0.878 & 58.0 & 0.997 & 591.2 & 0.758 & 24.1 \\
 & N & 0.069 & 6.5 & 0.000 & 0.0 & 0.000 & 0.0 & 0.001 & 0.0 & 0.405 & 3.1 & 0.893 & 426.8 & 0.870 & 210.0 \\
 & T & 0.098 & 7.1 & 0.016 & 0.1 & 0.009 & 0.1 & 0.003 & 0.0 & 0.965 & 133.3 & 1.000 & 8957.1 & 0.991 & 256.9 \\
 \midrule
\multirow{5}{*}{Assemb} & M & 0.077 & 30.5 & 0.023 & 0.3 & 0.000 & 0.0 & 0.004 & 0.4 & 0.457 & 30.6 & 1.000 & 8251.1 & 0.766 & 70.5 \\
 & D & 0.003 & 0.0 & 0.005 & 0.1 & 0.006 & 0.3 & 0.017 & 0.5 & 0.158 & 6.1 & 0.834 & 181.6 & 0.000 & 0.0 \\
 & O & 0.002 & 0.1 & 0.000 & 0.0 & 0.000 & 0.0 & 0.000 & 0.0 & 0.281 & 21.3 & 0.754 & 186.7 & 0.866 & 177.7 \\
 & N & 0.219 & 9.4 & 0.016 & 0.3 & 0.000 & 0.0 & 0.009 & 0.2 & 0.073 & 1.2 & 0.497 & 82.2 & 0.457 & 49.3 \\
 & T & 0.257 & 39.9 & 0.043 & 0.7 & 0.006 & 0.3 & 0.027 & 1.1 & 0.544 & 59.3 & 1.000 & 8701.5 & 0.987 & 297.5 \\
 \midrule
\multirow{5}{*}{X86-Sok} & M & 0.245 & 102.2 & 0.005 & 0.0 & 0.002 & 0.0 & 0.003 & 0.1 & 0.804 & 106.0 & 1.000 & 7414.6 & 0.696 & 51.5 \\
 & D & 0.004 & 0.1 & 0.001 & 0.0 & 0.000 & 0.0 & 0.035 & 0.4 & 0.336 & 33.1 & 0.956 & 334.6 & 0.000 & 0.0 \\
 & O & 0.004 & 0.1 & 0.000 & 0.0 & 0.000 & 0.0 & 0.000 & 0.0 & 0.660 & 72.9 & 0.883 & 447.0 & 0.486 & 44.6 \\
 & N & 0.780 & 538.5 & 0.003 & 0.0 & 0.000 & 0.0 & 0.007 & 0.1 & 0.222 & 4.0 & 0.848 & 612.9 & 0.763 & 250.9 \\
 & T & 0.818 & 640.9 & 0.008 & 0.0 & 0.002 & 0.0 & 0.045 & 0.5 & 0.895 & 216.0 & 1.000 & 8809.1 & 0.946 & 347.0 \\
 \midrule
\multirow{5}{*}{RW} & M & 0.266 & 65.8 & 0.004 & 0.0 & 0.015 & 0.1 & 0.000 & 0.0 & 0.831 & 345.6 & 0.990 & 6903.4 & 0.814 & 73.4 \\
 & D & 0.008 & 0.8 & 0.000 & 0.0 & 0.002 & 0.0 & 0.000 & 0.0 & 0.546 & 33.8 & 0.863 & 550.3 & 0.000 & 0.0 \\
 & O & 0.013 & 6.2 & 0.000 & 0.0 & 0.000 & 0.0 & 0.000 & 0.0 & 0.755 & 81.9 & 0.865 & 213.1 & 0.608 & 45.7 \\
 & N & 0.671 & 487.2 & 0.000 & 0.0 & 0.000 & 0.0 & 0.015 & 0.1 & 0.314 & 4.1 & 0.825 & 484.9 & 0.817 & 228.4 \\
 & T & 0.698 & 560.1 & 0.004 & 0.0 & 0.017 & 0.1 & 0.015 & 0.1 & 0.875 & 465.4 & 0.990 & 8151.7 & 0.951 & 347.5 \\
 \midrule
\multirow{5}{*}{Obf-Ben} & M & 1.000 & 12918.9 & 1.000 & 17.2 & 1.000 & 456.4 & 1.000 & 41.1 & 1.000 & 3879.5 & 1.000 & 23003.2 & 1.000 & 2487.0 \\
 & D & 1.000 & 2129.7 & 1.000 & 50.7 & 1.000 & 488.8 & 1.000 & 186.6 & 1.000 & 330.8 & 1.000 & 2643.0 & 0.000 & 0.0 \\
 & O & 0.727 & 2.3 & 0.000 & 0.0 & 0.000 & 0.0 & 0.000 & 0.0 & 1.000 & 1074.1 & 1.000 & 1241.8 & 1.000 & 430.3 \\
 & N & 0.000 & 0.0 & 0.000 & 0.0 & 0.636 & 0.8 & 0.800 & 1.4 & 1.000 & 6.2 & 1.000 & 72.9 & 1.000 & 3.8 \\
 & T & 1.000 & 15050.9 & 1.000 & 67.9 & 1.000 & 946.0 & 1.000 & 229.0 & 1.000 & 5290.6 & 1.000 & 26960.8 & 1.000 & 2921.1 \\
 \midrule
\multirow{5}{*}{Quarks} & M & 0.288 & 2.8 & 0.299 & 2.6 & 0.301 & 4.9 & 0.211 & 0.9 & 0.999 & 489.6 & 1.000 & 3448.9 & 0.959 & 332.0 \\
 & D & 0.127 & 1.1 & 0.130 & 1.3 & 0.163 & 1.4 & 0.053 & 0.2 & 0.724 & 35.2 & 1.000 & 1753.9 & 0.000 & 0.0 \\
 & O & 0.002 & 0.0 & 0.002 & 0.0 & 0.004 & 0.0 & 0.002 & 0.0 & 0.979 & 134.8 & 0.999 & 339.1 & 0.862 & 104.9 \\
 & N & 0.028 & 0.1 & 0.030 & 0.1 & 0.022 & 0.2 & 0.002 & 0.0 & 0.481 & 4.4 & 0.791 & 815.8 & 1.000 & 317.1 \\
 & T & 0.324 & 4.0 & 0.334 & 4.0 & 0.327 & 6.5 & 0.246 & 1.1 & 1.000 & 663.9 & 1.000 & 6357.7 & 1.000 & 753.9 \\
\bottomrule
\end{tabular}\vspace{-9pt}
\end{table*}

\noindent\textbf{Baselines.}
We compare our approach against several state-of-the-art rule-based and neural disassemblers. For rule-based systems, we include \textit{IDA Pro (v9.1)}~\cite{hex-raysIDAPro2025}, a widely adopted commercial tool, from which we programmatically extracted instruction addresses using its SDK (idalib). We also evaluated \textit{Ghidra (v11.3.2)}~\cite{nationalsecurityagencyGhidra2025}, a popular open-source suite, using pyghidra to dump instruction addresses. Additionally, we benchmarked against \textit{ddisasm (v1.9.0, commit 11d6ba92)}~\cite{flores-montoyaDatalogDisassembly2020a}, a research disassembler based on Datalog; this version has already incorporated improvements post-dating their Weighted Interval Scheduling (WIS) work~\cite{flores-montoyaDisassemblyWeightedInterval2025a}, and we parsed its exported GTIRB files. Our comparison with neural network-based disassemblers includes two key systems. We evaluated \textit{XDA}~\cite{Pei2020XDAAR} by fine-tuning the pre-trained model provided by its authors on the instruction boundary classification task for 30 epochs, adhering to their original paper's setup. We also included \textit{DeepDi}~\cite{Yu2022DeepDiLA}, for which we utilized the official model due to the absence of a publicly available implementation.

\noindent\textbf{Evaluation Metrics.} To evaluate the performance of our approach, we employ several metrics to assess both instruction-level accuracy and the frequency of constraint violations. For characterizing instruction-level accuracy, we utilize standard metrics: \textit{Precision}, \textit{Recall}, and the \textit{F1 score}. These are calculated based on the reference labels and provide a comprehensive measure of how accurately individual instructions are identified. To quantify the frequency of constraint violations, we use two distinct metrics: \textit{File-Level Error Rate}: This metric represents the proportion of files that contain at least one constraint violation relative to the total number of files analyzed. It provides an overview of how widespread errors are across the disassembly results. \textit{Byte-Normalized Error Rate}: To account for the varying sizes of binary files, this metric characterizes errors by dividing the total number of detected errors by the total number of bytes in the code section of all analyzed files. This normalization allows for a more fine-grained comparison of error propensity, irrespective of file size. To make the number more readable, we report the average errors per 1MB$(1024 \times 1024)$, instead of per byte.

\subsection{Violations of Constraints}

This section evaluates the prevalence of constraint violations in dataset labels and disassembler outputs. Since our pruning algorithm eliminates all of the violations, we report the error rate before pruning. While our error detection algorithm can operate without boolean labels, we implemented a score-based approach to align with our pruning algorithm. For rule-based disassemblers and dataset labels, we assign a score of $1$ to addresses labeled as instructions (true) and $-1$ to those labeled as non-instructions (false). For binary classification models, including ours and DeepDi, we use their output logits as scores. For XDA, a multi-class classification model, we use the probability of an address being an instruction start and apply an inverse sigmoid function to derive the score.

\begin{table}[!tbp]
\vspace{-5pt}
\centering
\caption{Errors Detected in Dataset Labels.}
\label{tab:error_labels}
\sisetup{
  group-separator={,}, group-minimum-digits = 4 }
\setlength{\tabcolsep}{4pt}
\small
\resizebox{\linewidth}{!}{\begin{tabular}{@{}c S[table-format=4.0]  S[table-format=6.0]  S[table-format=6.0]  S[table-format=6.0]  S[table-format=6.0]  S[table-format=5.0]@{}} \toprule
Stats & {Pangine} & {Assemb} & {X86-Sok} & {RW} & {Obf-Ben} & {Quarks} \\
\midrule
M         & 391   & 294670 & 135156 & 14801  & 160287 & 2825   \\
D         & 0     & 351    & 85     & 185    & 26424  & 1122   \\
O          & 12    & 689    & 192    & 1397   & 28     & 4      \\
N         & 4423  & 90760  & 712429 & 109577 & 0      & 133    \\
T       & 4826  & 386470 & 847862 & 125960 & 186739 & 4084   \\
Files       & 879   & 81875  & 3997   & 526    & 11     & 1298   \\
\bottomrule
\end{tabular}
}
\vspace{-10pt}
\end{table}
 
Table~\ref{tab:error} presents the evaluation results for constraint violations, where \textbf{M} denotes \textit{Missing Post-Dominator} (MPD), \textbf{D} signifies \textit{Dead-End Sequence} (DES), and \textbf{O} represents \textit{Overlapping Instructions} (OI). A significant portion of detected MPD errors involve instructions post-dominated by a \texttt{NOP} instruction not recognized as code. These instances often indicate missing padding bytes within a basic block or mark a virtual exit and are effectively dead code. We list these as \textbf{N} (NOP), separate them from other, more critical, MPD violations. \textbf{T} denotes total of all errors. The detailed statistics for error counts in dataset labels are shown in Table~\ref{tab:error_labels}.

Several key conclusions emerge from these results:

\noindent\textbf{All dataset labels contain constraint violations.} This finding aligns with previous research~\cite{flores-montoyaDisassemblyWeightedInterval2025a}, where errors were manually identified. Our approach systematically locates these violations without requiring manual intervention or another referencing tool's results. The errors often have common patterns, such as missing \texttt{NOP}s within basic blocks or missing jump targets, indicating false negatives. The errors not belonging to any such patterns usually indicate false positives. These findings can help identify bugs in the dataset construction scripts, enhancing the quality of the resulting datasets. Tady has more \texttt{NOP} related errors but fewer errors of other types than DeepDi. After investigation, we found that its training set Pangine wrongly labels many \texttt{NOP}s within basic blocks as false, resulting in the wrong behavior of the model. This shows the necessity of high-quality training sets. 

\noindent\textbf{Rule-based disassemblers (IDA, Ghidra, ddisasm) still exhibit errors.} Although these tools are designed to enforce constraints, leading to relatively better performance, violations were identified in their outputs. Manual investigation verified these as genuine disassembler errors. This finding indicates that they might be tricked into invalid intermediate states without effective rollback mechanisms, particularly when encountering obfuscated code.

\noindent\textbf{Neural disassemblers struggle to enforce consistency through architectural design alone.} For instance, DeepDi does not prevent all overlapping instructions though utilizing overlapping edges in its design. Our model, Tady, successfully prevents DES but does not fully mitigate MPD and OI issues, this highlights the importance of a rule based regularization upon neural disassemblers' outputs. 

\noindent\textbf{Obfuscation significantly increases consistency violations.} Code from the Obf-Benchmark and Quarks datasets caused all disassemblers to generate substantially more violations compared to other datasets. This indicates that the obfuscation techniques effectively confused the disassemblers, leading them to incomplete or incorrect intermediate states. This observation offers insight into why even rule-based disassemblers produce constraint violations.

\subsection{Accuracy Evaluation}

In this section, we evaluate the instruction-level performance of various disassemblers using labels provided by the datasets. While these dataset labels may contain errors, meaning absolute performance cannot be precisely determined, the relative performance metrics remain valuable for comparative analysis of the disassemblers.

We evaluated the disassemblers on both common and obfuscated binaries. For this experiment, in addition to the previously introduced \textit{Tady} model (trained on the Pangine dataset), we constructed another model, named \textit{TadyA}, which was trained for one epoch on the training portion of a composite dataset created from all available datasets to simulate a train-validation split. Specifically, this composite dataset was generated by sampling 2 files from obf-benchmark and 100 files from each of the other five datasets. We then performed a 9:1 train-validation split on this mixed dataset.

\begin{table}[t!]
\vspace{-5pt} \centering
\caption{Comparison of Precision, Recall, and F1 Before and After Pruning for Various Datasets and Disassemblers.}
\label{tab:pruning_comparison_wide}
\sisetup{
  round-mode=places,
  round-precision=4, table-number-alignment=center }
\setlength{\tabcolsep}{1pt} \small \resizebox{\linewidth}{!}{\begin{tabular}{@{}ccc S[table-format=1.4] S[table-format=1.4] S[table-format=1.4] S[table-format=1.4] S[table-format=1.4] S[table-format=1.4] S[table-format=1.4]@{}} \toprule
Dataset & Metric & State & {IDA} & {Ghidra} & {Ddisasm} & {DeepDi} & {XDA} & {TadyA} & {Tady} \\
\midrule
\multirow{6}{*}{Pangine} & \multirow{2}{*}{P} & B & 0.9999914 & {\bfseries 1.0000} & 0.9999832 & 0.9984852 & 0.9612533 & 0.9987902 & 0.9996224 \\
&                    & A  & {\bfseries 1.0000} & 0.9999319 & 0.9999557 & 0.9990888 & 0.9893466 & 0.9994082 & 0.9998850 \\
\cmidrule(lr){2-10}
& \multirow{2}{*}{R} & B & 0.9884336 & 0.9331979 & 0.9995879 & 0.9995256 & 0.9630268 & 0.9994173 & {\bfseries 0.9999} \\
&                    & A  & 0.9884333 & 0.9331952 & 0.9995869 & 0.9998182 & 0.9906347 & 0.9990309 & {\bfseries 0.9999} \\
\cmidrule(lr){2-10}
& \multirow{2}{*}{F1} & B & 0.9941789 & 0.9654442 & {\bfseries 0.9998} & 0.9990051 & 0.9621392 & 0.9991037 &  0.9997474 \\
&                    & A  & 0.9941742 & 0.9654116 & 0.9997712 & 0.9994533 & 0.9899903 & 0.9992196 & {\bfseries 0.9999} \\
\midrule
\multirow{6}{*}{Assemb} & \multirow{2}{*}{P} & B & 0.9959019 & {\bfseries 0.9999} & 0.9995734 & 0.9989723 & 0.9747387 & 0.9964502 & 0.9980094 \\
&                    & A  & 0.9961671 & {\bfseries 0.9999} & 0.9995770 & 0.9996769 & 0.9897426 & 0.9979152 & 0.9992380 \\
\cmidrule(lr){2-10}
& \multirow{2}{*}{R} & B & 0.9858763 & 0.9558540 & 0.9914588 & 0.9954365 & 0.9251807 & {\bfseries 0.9956} & 0.9934821 \\
&                    & A  & 0.9863590 & 0.9561768 & 0.9915969 & {\bfseries 0.9956} & 0.9678894 & 0.9953210 & 0.9939360 \\
\cmidrule(lr){2-10}
& \multirow{2}{*}{F1} & B & 0.9908638 & 0.9773828 & 0.9954995 & {\bfseries 0.9972} & 0.9493133 & 0.9960245 & 0.9957406 \\
&                    & A  & 0.9912388 & 0.9775530 & 0.9955709 & {\bfseries 0.9976} & 0.9786940 & \bfseries 0.9966164 & 0.9965800 \\
\midrule
\multirow{6}{*}{X86-Sok} & \multirow{2}{*}{P} & B & {\bfseries 0.9955} & 0.9950230 & 0.9889478 & 0.9703897 & 0.9575027 & 0.9906794 & 0.9791626 \\
&                    & A  & {\bfseries 0.9954} & 0.9944379 & 0.9889725 & 0.9716127 & 0.9795238 & 0.9909482 & 0.9790827 \\
\cmidrule(lr){2-10}
& \multirow{2}{*}{R} & B & 0.9860424 & 0.9773080 & {\bfseries 0.9999} & 0.9994387 & 0.9584353 & 0.9995672 & 0.9989376 \\
&                    & A  & 0.9859360 & 0.9787640 & {\bfseries 0.9999} & 0.9984926 & 0.9917596 & 0.9996444 & 0.9990589 \\
\cmidrule(lr){2-10}
& \multirow{2}{*}{F1} & B & 0.9907503 & 0.9860860 & 0.9944054 & 0.9847000 & 0.9579688 & {\bfseries 0.9951} & 0.9889512 \\
&                    & A  & 0.9906416 & 0.9865387 & 0.9944176 & 0.9848692 & 0.9856037 &  {\bfseries 0.9953} & 0.9889699 \\
\midrule
\multirow{6}{*}{RW} & \multirow{2}{*}{P} & B & {\bfseries 0.9963} & 0.9957687 & 0.9855553 & 0.9767243 & 0.9531378 & 0.9925366 & 0.9843275 \\
&                    & A  & {\bfseries 0.9963} & 0.9957698 & 0.9855532 & 0.9786137 & 0.9775635 & 0.9925352 & 0.9847857 \\
\cmidrule(lr){2-10}
& \multirow{2}{*}{R} & B & 0.9907754 & 0.9226709 & {\bfseries 0.9998} & 0.9980865 & 0.9533159 & 0.9994754 & 0.9905881 \\
&                    & A  & 0.9907714 & 0.9228141 & {\bfseries 0.9998} & 0.9838108 & 0.9910771 & 0.9996220 & 0.9906344 \\
\cmidrule(lr){2-10}
& \multirow{2}{*}{F1} & B & 0.9935527 & 0.9578272 & 0.9926170 & 0.9872899 & 0.9532268 & {\bfseries 0.9960} & 0.9874479 \\
&                    & A  & 0.9935361 & 0.9579049 & 0.9926156 & 0.9812054 & 0.9842739 & {\bfseries 0.9961} & 0.9877014 \\
\midrule
\multirow{6}{*}{Obf-Ben} & \multirow{2}{*}{P} & B & 0.8451757 & 0.6827769 & 0.6945186 & 0.8643398 & 0.6640892 & {\bfseries 0.9237} & 0.9039955 \\
&                    & A  & 0.8673666 & 0.7105717 & 0.7063917 & 0.8702054 & 0.8053187 & {\bfseries 0.9008 }& 0.8934074 \\
\cmidrule(lr){2-10}
& \multirow{2}{*}{R} & B & 0.0911837 & 0.4464715 & 0.1549890 & 0.9316564 & 0.8005244 & {\bfseries 0.9783} & 0.9624078 \\
&                    & A  & 0.0893659 & 0.4323975 & 0.1514236 & 0.8930388 & 0.8220970 & {\bfseries 0.9278} & 0.9197217 \\
\cmidrule(lr){2-10}
& \multirow{2}{*}{F1} & B & 0.1646082 & 0.5398997 & 0.2534239 & 0.8967365 & 0.7259520 & {\bfseries 0.9502} & 0.9322876 \\
&                    & A  & 0.1620369 & 0.5376338 & 0.2493879 & 0.8814743 & 0.8136214 & {\bfseries 0.9141} & 0.9063736 \\
\midrule
\multirow{6}{*}{Quarks} & \multirow{2}{*}{P} & B & {\bfseries 1.0000} & 0.9981754 & 0.9983600 & 0.9948031 & 0.9436001 & 0.9974530 & 0.9952361 \\
&                    & A  &  {\bfseries 1.0000} & 0.9990393 & 0.9983457 & 0.9971462 & 0.9848929 & 0.9977614 & 0.9965926 \\
\cmidrule(lr){2-10}
& \multirow{2}{*}{R} & B &  {\bfseries 1.0000} & 0.9190927 & 0.9981868 & 0.9971532 & 0.9699871 & 0.9994154 & 0.9958163 \\
&                    & A  & {\bfseries 0.9999}  & 0.9433250 & 0.9981027 & 0.9880242 & 0.9945844 & 0.9995055 & 0.9974448 \\
\cmidrule(lr){2-10}
& \multirow{2}{*}{F1} & B &  {\bfseries 1.0000}  & 0.9570031 & 0.9982733 & 0.9959768 & 0.9566117 & 0.9984332 & 0.9955261 \\
&                    & A  &  {\bfseries 0.9999} & 0.9703831 & 0.9982242 & 0.9925642 & 0.9897149 & 0.9986327 & 0.9970185 \\
\bottomrule
\end{tabular}}\end{table}
 
\noindent\textbf{Overall Performance.} As shown in Table~\ref{tab:pruning_comparison_wide}, neural disassemblers, despite potentially lower precision, usually outperform rule-based disassemblers in terms of recall. Our \textit{TadyA} model, trained on a small portion of the combined datasets, generalized well to the remaining data, often yielding the best performance among all evaluated disassemblers, achieving the best performance on three of the six benchmarks and performs competently on others. Furthermore, our \textit{Tady} model, trained exclusively on the Pangine dataset, demonstrated robust performance on other datasets featuring unseen compilers and binaries, consistently outperforming its direct competitor, DeepDi, on all Linux benchmarks.

\noindent\textbf{Pruning.} In most cases, the pruning algorithm has a positive impact on the F1 score. This is particularly evident for XDA, which assigns appropriate weights in general but does not explicitly considers the interdependence between instructions. For other disassemblers, either internal nodes are recalled as positive or the successors of negative-weighted nodes are pruned. How the F1 score changes depends on the quality of the assigned weights. Since the purpose of the pruning algorithm is to enforce constraints, an improvement in accuracy is not guaranteed, though often observed. 

\noindent\textbf{Anti-Obfuscation.}
Notably, despite no prior training on obfuscated data, \textit{Tady} exhibited significant robustness against anti-disassembly obfuscators in the Obf-Benchmark dataset. It outperformed rule-based disassemblers by a large margin; for instance, ddisasm and IDA identified only a few instructions. Ghidra, while achieving better results than IDA Pro and ddisasm, required over an hour to disassemble these binaries. This finding aligns well with DeepDi's experiments~\cite{Yu2022DeepDiLA}.

\noindent\textbf{Indirect Jumps.}
The robustness of our disassembler against indirect jumps was demonstrated with its performance against obfuscators. The Tigress obfuscator, employed in the Quarks dataset, implements several techniques that introduce complex indirect jumps, including \textit{virtualization}, \textit{Control Flow Flattening} (CFF), \textit{mix1}, and \textit{mix2}.
To confirm that these obfuscations indeed introduce indirect jumps, we compared the ratio of indirect jumps in binaries before and after applying obfuscation. Our analysis revealed that indirect jumps were at least five times more frequent post-obfuscation. Specifically, CFF applied at the O0 optimization level increased the ratio of indirect jumps by a factor of $17.32$. \textit{Tady}'s accuracy remained high in these scenarios, achieving F1 scores above $0.999$, which demonstrates its robustness against indirect jumps.

\subsection{Efficiency Evaluation}

This section evaluates the efficiency of our proposed disassembler, \textit{Tady}, including the model and the subsequent pruning algorithm. Our evaluation consists of time/memory consumption, and a detailed breakdown of computational costs.

To create a representative dataset, we first sorted binaries in x86-sok by their code section size. We then sampled binaries by selecting from exponentially increasing size intervals, choosing up to three samples from each interval where available. The disassembly script, previously employed in our analyses, was run in headless mode. We recorded the execution time for each disassembler to facilitate a comparative efficiency analysis. Rule-based disassemblers were benchmarked on a single Core i9-12900K CPU, while neural network-based disassemblers utilized an NVIDIA A6000 Ada GPU.

\begin{figure}[ht]
    \centering
    \includegraphics[width=\linewidth]{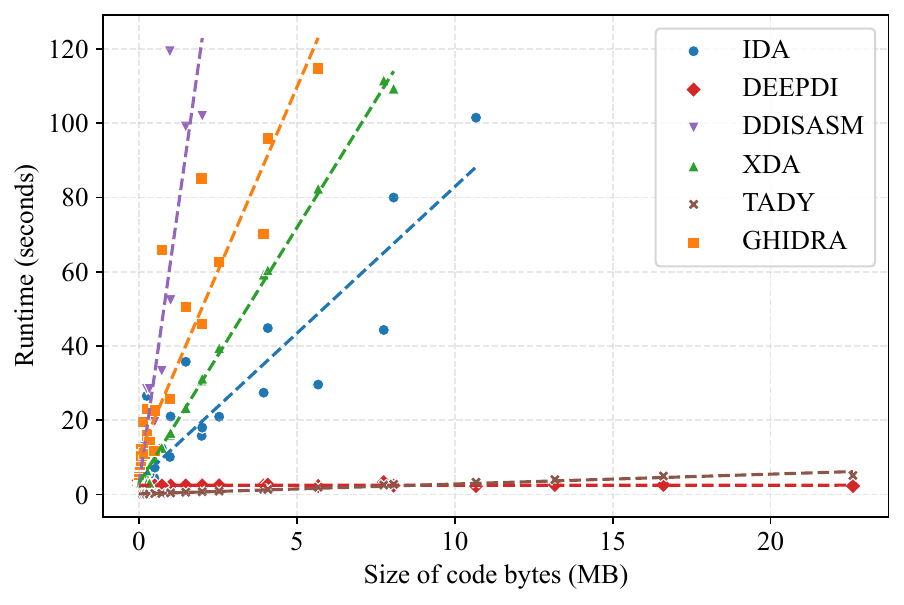}
    \vspace{-20pt} \caption{Run time comparison of the disassemblers.}
    \label{fig:efficiency}
\end{figure}

\noindent\textbf{Disassembly Efficiency.} Figure~\ref{fig:efficiency} shows the relationship between code section size and disassembly time for various disassemblers. The results show that \textit{Tady} achieves the second-fastest performance, surpassed only by DeepDi. Notably, \textit{Tady} is significantly faster than the widely-used rule-based disassembler, IDA Pro.

\begin{figure}[ht]
    \centering
    \includegraphics[width=\linewidth]{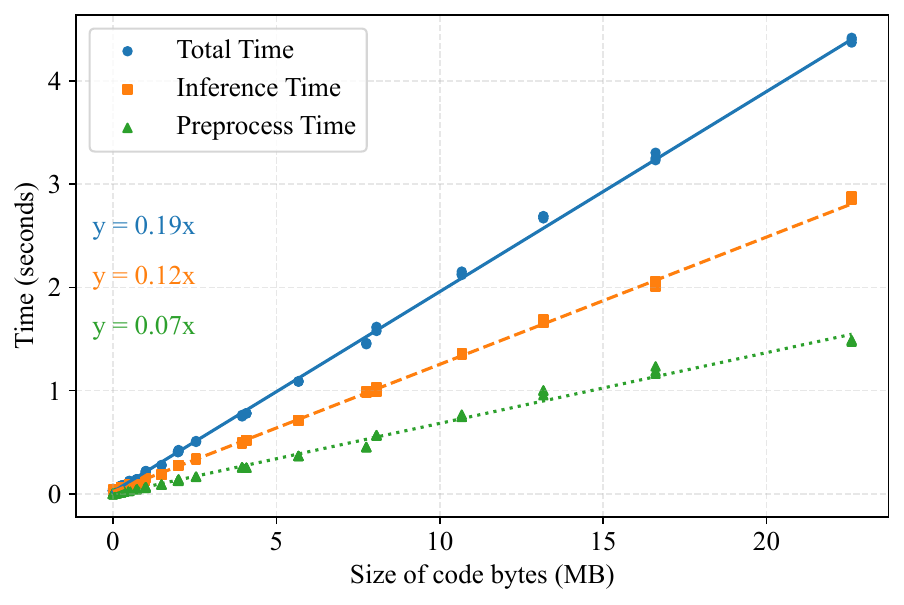}
    \vspace{-20pt} \caption{Time Consumption of Disassemble runtime.}
    \label{fig:time_composition}
\end{figure}

To further quantify \textit{Tady}'s efficiency and analyze its time consumption components, we conducted a rigorous benchmark. This involved recording a detailed breakdown of the time consumed by each disassembly step. These steps include the \textit{Preprocessing} phase, which involves superset disassembly executed on a single CPU, and the \textit{Model Inference} phase, which is performed on the GPU with the batch size and sequence length empirically set to 32 and 8192, respectively.

As depicted in Figure~\ref{fig:time_composition}, \textit{Tady}'s total throughput is 5.26 MB/s. The CPU-bound preprocessing step accounts for a significant portion (37\%) of the total time. Despite this, \textit{Tady}'s overall performance remains considerably faster than IDA Pro, highlighting its substantial efficiency and practical applicability for large-scale binary analysis.

\begin{figure}[!t]
    \centering
    \includegraphics[width=\linewidth]{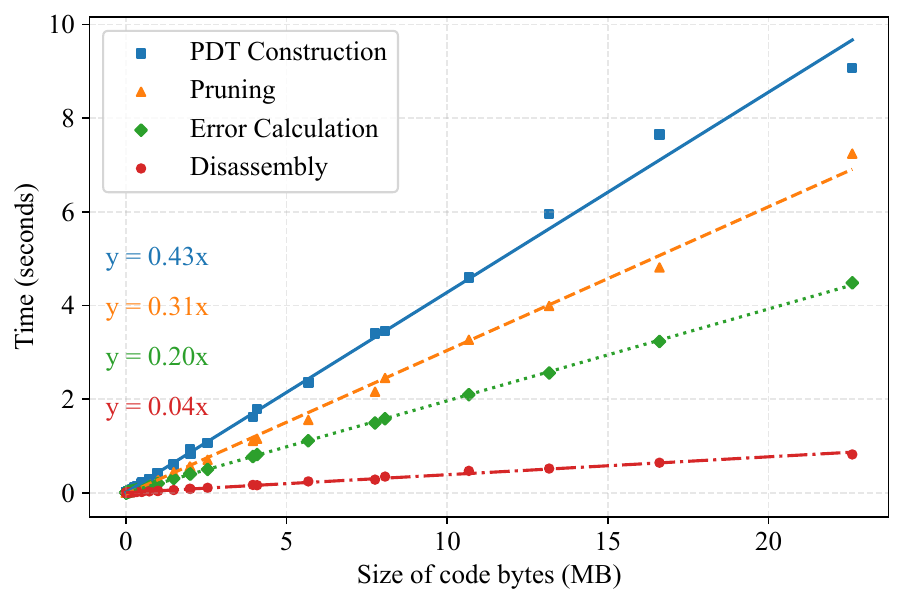}
    \vspace{-20pt} \caption{Time Consumption of the Algorithms.}
    \label{fig:prune_time}
\end{figure}

Beyond the disassembly phase including the preprocessing and model inference, we also evaluated the time and memory efficiency of our error detection and pruning algorithm.

\noindent\textbf{Time Consumption.}
Figure~\ref{fig:prune_time} shows how the algorithms' execution time correlates with the binaries' code section size. The processing time exhibits a linear relationship with code size, which aligns perfectly with our theoretical analysis predicting linear time complexity for these algorithms in Appendix~\ref{sec:complexity}. The pruning components are also efficient, achieving throughputs of 2.33 MB/s for the PDT construction process and 3.23 MB/s for the pruning algorithm itself. Even when factoring in the time for these post-processing steps, \textit{Tady} maintains its high overall speed.

\begin{figure}[!t]
    \centering
    \includegraphics[width=\linewidth]{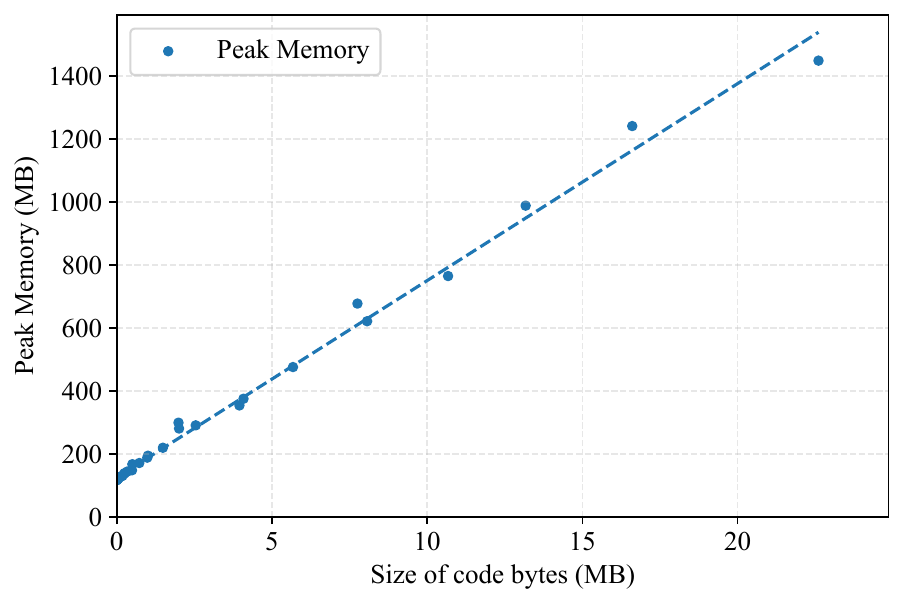}
    \vspace{-20pt} \caption{Peak Memory Usage for PDT Construction.}
    \label{fig:prune_memory}
\end{figure}

\noindent\textbf{Memory Usage.} Figure~\ref{fig:prune_memory} shows the linear relationship between peak memory usage and the size of the code section. Processing the PDT at the WCC level effectively reduces memory requirements for tree construction. Since most WCCs are concentrated within a comparatively small size range, memory usage is primarily dominated by storing edges. Consequently, the memory needed to process individual WCCs is comparatively small. This finding underscores the effectiveness of our approach in managing WCC size by strategically omitting call edges during their construction.

\subsection{Ablation Study}
To evaluate the impact of our core design choices, we conducted an ablation study focusing on \textbf{masked sliding window attention (MSWA)} and \textbf{message passing (MP)}. We assessed their effects on accuracy and constraint violations by training four model variants with different combinations of these mechanisms on the same dataset. Performance was evaluated on our test datasets (Tables~\ref{tab:ablation_error} and~\ref{tab:ablation_prf}).

\begin{table}[!ht]
\centering
\caption{Ablation Study: Comparison of Model Configurations Across Datasets and Error Metrics (F: Total File Error Rate, B: Total Errors per 1MB).}
\label{tab:ablation_error}
\sisetup{
  round-mode=places,
  table-align-exponent = false,
  output-exponent-marker = \text{e}
}
\setlength{\tabcolsep}{3pt}
\small
\resizebox{\linewidth}{!}{\begin{tabular}{@{}l
                S[table-format=1.3, round-precision=3] S[table-format=5.1, round-precision=1] S[table-format=1.3, round-precision=3] S[table-format=5.1, round-precision=1] S[table-format=1.3, round-precision=3] S[table-format=5.1, round-precision=1] S[table-format=1.3, round-precision=3] S[table-format=5.1, round-precision=1]@{}} \toprule
\multirow{2}{*}{Dataset} & \multicolumn{2}{c}{MSWA + MP} & \multicolumn{2}{c}{MSWA} & \multicolumn{2}{c}{SWA + MP} & \multicolumn{2}{c}{SWA} \\
\cmidrule(lr){2-3} \cmidrule(lr){4-5} \cmidrule(lr){6-7} \cmidrule(lr){8-9}
 & {F} & {B} & {F} & {B} & {F} & {B} & {F} & {B} \\
\midrule
X86-Sok   & 0.946 & 347.0  & 0.924 & 337.9  & 1.000 & 12326.9 & 1.000 & 2528.4 \\
Pangine       & 0.991 & 256.9  & 0.973 & 223.6  & 1.000 & 14180.8 & 1.000 & 1639.7 \\
RW            & 0.951 & 347.5  & 0.899 & 352.2  & 1.000 & 11970.7 & 0.990 & 1954.9 \\
Assemb    & 0.987 & 297.5  & 0.999 & 267.5  & 1.000 & 17479.5 & 1.000 & 2587.1 \\
Obf-Ben & 1.000 & 2921.1 & 1.000 & 3779.7 & 1.000 & 17398.7 & 1.000 & 12524.6 \\
Quarks        & 1.000 & 753.9  & 1.000 & 499.2  & 1.000 & 14566.6 & 1.000 & 3456.8 \\
\bottomrule
\end{tabular}}\end{table}
 \begin{table}[!ht]
\vspace{-5pt} \centering
\caption{Ablation Study: Comparison of Model Configurations Across Datasets (Precision, Recall, F1-Score - Pruned).}
\label{tab:ablation_prf}
\sisetup{
  round-mode=places,
  round-precision=3, table-align-exponent = false,
  output-exponent-marker = \text{e}
}
\setlength{\tabcolsep}{2.5pt} \resizebox{\linewidth}{!}{\begin{tabular}{@{}l
                S[table-format=0.3] S[table-format=0.3] S[table-format=0.3] S[table-format=0.3] S[table-format=0.3] S[table-format=0.3] S[table-format=0.3] S[table-format=0.3] S[table-format=0.3] S[table-format=0.3] S[table-format=0.3] S[table-format=0.3]@{}} \toprule
\multirow{2}{*}{Dataset} & \multicolumn{3}{c}{MSWA + MP} & \multicolumn{3}{c}{MSWA} & \multicolumn{3}{c}{SWA + MP} & \multicolumn{3}{c}{SWA} \\
\cmidrule(lr){2-4} \cmidrule(lr){5-7} \cmidrule(lr){8-10} \cmidrule(lr){11-13}
 & {P} & {R} & {F1} & {P} & {R} & {F1} & {P} & {R} & {F1} & {P} & {R} & {F1} \\
\midrule
X86-Sok
& 0.979 & 0.999 & 0.989 & 0.977 & 0.999 & 0.988 & 0.957 & 0.943 & 0.950 & 0.972 & 0.998 & 0.985 \\
Pangine
& 1.000 & 1.000 & 1.000 & 1.000 & 1.000 & 1.000 & 0.977 & 0.956 & 0.966 & 0.997 & 0.999 & 0.998 \\
RW
& 0.985 & 0.991 & 0.988 & 0.982 & 0.991 & 0.987 & 0.965 & 0.936 & 0.950 & 0.978 & 0.991 & 0.984 \\
Assemb
& 0.999 & 0.994 & 0.997 & 0.999 & 0.994 & 0.996 & 0.961 & 0.914 & 0.937 & 0.991 & 0.988 & 0.990 \\
Obf-Ben
& 0.893 & 0.920 & 0.906 & 0.857 & 0.917 & 0.886 & 0.895 & 0.855 & 0.875 & 0.849 & 0.914 & 0.880 \\
Quarks
& 0.997 & 0.997 & 0.997 & 0.996 & 0.999 & 0.997 & 0.951 & 0.863 & 0.905 & 0.985 & 0.995 & 0.990 \\
\bottomrule
\end{tabular}}\vspace{-10pt}
\end{table}
 
The variants, each trained for one epoch on the Pangine dataset, were:
(1) Full Model: MSWA + MP; (2) MSWA only; (3) SWA + MP: Standard sliding window attention~(SWA) with MP; (4) SWA only (Baseline). SWA utilizes a full attention mask within the sliding window, contrasting with MSWA's specialized reachability masks.

Results in Table~\ref{tab:ablation_error} and Table~\ref{tab:ablation_prf} show that \textbf{MSWA} significantly outperforms SWA, consistently yielding higher accuracy and fewer constraint violations. This indicates our reachability attention mask effectively injects execution semantics, enhancing performance.
The \textbf{MP} mechanism's impact was conditional: detrimental with SWA but beneficial with MSWA, particularly improving accuracy on the obfuscation dataset while largely preserving performance elsewhere. This supports the idea that while local context often suffices for disassembly, global information from MP is advantageous for complex cases like obfuscated code.

 \section{Discussion}

Our pruning and error detection algorithm assumes an instruction with fall-through semantics is immediately post-dominated by its subsequent instruction. This generally holds, except for signal-based hardware exceptions (e.g., segmentation faults, division-by-zero errors). If an instruction triggers such an exception, control may transfer to an exception handler rather than the next instruction in memory. This scenario, particularly when exceptions are deliberately engineered for obfuscation as seen in techniques like~\cite{Popov2007BinaryOU}, represents a current limitation of Tady. High-level language exceptions such as try-catch blocks typically use mechanisms like function calls and do not violate this post-domination assumption. \section{Related Work}
\noindent\textbf{Regularizing Disassembly Results.} Previous works have explored regularizing disassembly output using constraints. Pdisasm~\cite{Miller2019ProbabilisticD} used an iterative algorithm with control-flow and data-flow features. Ddisasm~\cite{flores-montoyaDatalogDisassembly2020a, flores-montoyaDisassemblyWeightedInterval2025a} employed logic programming and later modeled disassembly as a weighted interval scheduling problem. D-Arm~\cite{Ye2023DARMDA} used data-flow logic, framing it as a maximum weight independent set problem solved with approximate algorithms. These approaches often assign weights and solve optimization problems but can struggle with the complexity of intricate constraints. Our work aligns with this trend but introduces the post-dominator tree as a novel backbone for efficient error detection and constraint enforcement, addressing this complexity.

\noindent\textbf{Neural Disassemblers.} Neural network-based disassemblers aim to reduce manual rule creation through data-driven learning. Early efforts like ByteWeight~\cite{baoBYTEWEIGHTLearningRecognize2014} focused on function boundary identification. Subsequent models, such as XDA~\cite{Pei2020XDAAR} using transformers and DeepDi~\cite{Yu2022DeepDiLA} employing Graph Neural Networks, advanced instruction classification from raw bytes. While promising, these neural approaches often produce outputs violating fundamental disassembly constraints. Our work specifically addresses these consistency issues to facilitate more reliable neural disassembly.

\noindent\textbf{Assembly Models.} Various neural architectures have been applied to binary analysis. Instruction2Vec~\cite{Lee2019Instruction2vecEP} and DeepVSA~\cite{guoDEEPVSAFacilitatingValueset2019} used word2vec and RNNs for instruction embeddings, while PalmTree~\cite{Li2021PalmTreeLA}, jTrans~\cite{Wang2022jTransJT}, and HermesSim~\cite{He2024CodeIN} utilized transformers or graph-based methods on intermediate representations. However, these models are unsuitable for our problem because they typically: (1)~presuppose already accurate disassembly with known instruction boundaries, but we need to identify valid instructions from a superset; (2)~process single, valid instruction sequences, whereas we must distinguish valid instructions among multiple potential traces; and (3)~are not designed to process multiple potential traces in parallel, which is required in our superset scenario.

\section{Conclusion}

In this work, we systematically regularize the solution space of the disassembly problem with the structural constraints derived from post-dominance relation and the non-overlapping instructions assumption. We propose an efficient error detection algorithm based on the post-dominator tree over the superset CFG. This error detection algorithm successfully identified various errors from neural disassemblers, rule-based disassemblers and those disassembly datasets' labels without relying on ground truth. We mitigate the constraint violations by providing a better model design, exploiting both local sequential feature and global graph structure of the problem. In addition, we provide a pruning mechanism to post-process the output of the model and completely eliminate the violations from the final disassembly result. Our proposed method can serve as a general post-processing step that enhances the usability of all neural network-based disassemblers.

\section*{Acknowledgments}

We would like to sincerely thank all the reviewers for their insightful feedback that greatly helped us to improve this paper. Additionally, special thanks are extended to Miaoqian Lin for her invaluable comments. The authors from  Institute of Information Engineering are supported in part by NSFC (U24A20236, 92270204) and CAS Project for Young Scientists in Basic Research (Grant No. YSBR-118). The authors from Tsinghua University are supported in part by NSFC (U24A20337) and the Joint Research Center for System Security, Tsinghua University (Institute for Network Sciences and Cyberspace) - Science City (Guangzhou) Digital Technology Group Co., Ltd.. \newpage
\section*{Ethical Considerations}

Our research was conducted with careful consideration of ethical implications, following the principles outlined in The Menlo Report and USENIX Security's ethical guidelines. We identified and analyzed potential impacts on all stakeholders, including end users, system administrators, and security practitioners. Our methodology prioritized minimizing risks while maximizing benefits to the security community. We maintained compliance with relevant terms of service and legal requirements throughout the study. We acknowledge the dual-use potential of our findings and have carefully balanced research transparency with potential misuse concerns.
 \section*{Open Science}

In accordance with USENIX Security's open science policy, the research artifacts associated with this paper are made publicly available at \url{https://doi.org/10.5281/zenodo.15541311}, including datasets, models, and source code. The latest version of the code will be maintained and updated at \url{https://github.com/5c4lar/tady}. 
\bibliographystyle{plain}
\bibliography{disasm.bib}

\section*{Appendix}
\appendix

\section{Complexity Analysis}
\label{sec:complexity}

Our pre-processing step involves a superset disassembly, where decoding occurs at every possible address. Although this step can be parallelized or GPU-accelerated, we observed it does not represent a computational bottleneck unless printable disassembly output is required. Decoding to the internal representation is efficient, even for large binaries, and exhibits linear time complexity, $O(L)$,  where $L$ is the size of the binary.

For our model, each instruction attends to a fixed-size local context (via a sliding window) and a fixed number of global connections. Consequently, the computational cost scales linearly with the sequence length, $O(S)$, where $S$ is the number of instructions in the sequence. This linear scaling enables the model to process long sequences effectively. The sliding window size determines the number of neighboring instructions to which each instruction attends, directly influencing the total computation in a linear fashion.

The PDT construction process comprises several key steps: WCC computation, SCC computation, and the calculation of immediate post-dominators within each WCC using the Lengauer-Tarjan algorithm.

Classical algorithms for WCC and SCC detection have a complexity of $O(N+E)$, where $N$ is the number of nodes (instructions) and $E$ is the number of edges in the Control Flow Graph CFG. In our specific CFGs, each node has at most two outgoing edges (for conditional branches), meaning $E \le 2N$. Therefore, the complexity for WCC and SCC computation simplifies to $O(N)$.

The Lengauer-Tarjan algorithm for computing immediate post-dominators exhibits a time complexity of $O(E \cdot \alpha(N))$, where $\alpha$ is the extremely slowly growing inverse Ackermann function~\cite{Lengauer1979AFA}. Given that $E \le 2N$ in our case, and $\alpha$ grows so slowly it is considered nearly constant for practical input sizes, this complexity is effectively linear.

Since we perform these computations within individual WCCs, and call edges are not connected during this phase (resulting in smaller WCCs), the actual runtime for PDT construction is determined by the size of the largest WCC and remains, in practice, linear with respect to the total number of nodes. Thus, the overall time complexity for PDT construction can be considered $O(N)$.

Following PDT construction, the error detection and pruning algorithms involve straightforward traversals of the PDT. Error detection requires a single pass. Pruning utilizes two passes: one for weight propagation up the tree and another for result collection down the tree. In a tree structure like the PDT, each node has at most one parent. Therefore, these traversals are also $O(N)$.

Based on the preceding analysis, the entire Tady workflow, including the error detection and pruning algorithms (as discussed in Section~\ref{sec: constraints}), demonstrates an overall linear time complexity and is expected to scale efficiently with the size of the input binaries.

\section{Evaluation on Commercial Obfuscator}
In addition to open-source obfuscators, we evaluated the disassemblers against a prominent commercial obfuscator, VMProtect~\cite{vmprotectsoftwareVMProtect2025}. This obfuscator supports several protection methods, including mutation, virtualization, or "Ultra" (a combination of the previous two). Mutation obfuscates the binary using instructions from the same Instruction Set Architecture (ISA). Virtualization, on the other hand, introduces a virtual machine and encodes the protected function into byte sequences specific to that virtual machine. Such potent obfuscation techniques, particularly virtualization, are beyond the scope of our current work. Consequently, our evaluation focused on the mutation capabilities of VMProtect. To simplify the evaluation and maintain a focus on static analysis, we disabled VMProtect's "Pack the output file" option. This option typically compresses the output binary, with decompression occurring only at runtime, effectively acting as an anti-static analysis measure. Analyzing such packed binaries is outside the scope of our current research. Our primary interest in this work is the analysis of statically obfuscated code.

Under this configuration, we obfuscated an example binary provided with the VMProtect Demo: \texttt{Licensing/BCB}. Recognizing that obfuscation patterns are often repetitive, we focused our efforts by obfuscating only the first protected function, \texttt{btTryClick}, and then manually analyzing it to establish the ground truth for this function. The results for this test case are presented below:

\begin{table}[htbp]
\centering
\caption{Performance on the VMProtect Example case.}
\label{tab:disassembler_prf}
\sisetup{
  round-mode=places,
  round-precision=3, table-number-alignment=center
}
\begin{tabular}{@{}l S[table-format=1.3]  S[table-format=1.3]  S[table-format=1.3]@{}} \toprule
Disassembler & {P} & {R} & {F1} \\
\midrule
IDA          & 1.000 & 0.013 & 0.026 \\
Ghidra       & 0.000 & 0.000 & 0.000 \\
DeepDi       & 1.000 & 1.000 & 1.000 \\
Ddisasm      & 1.000 & 1.000 & 1.000 \\
TadyA         & 1.000 & 1.000 & 1.000 \\
Tady         & 1.000 & 0.974 & 0.987 \\
Tady(pruned) & 1.000 & 1.000 & 1.000 \\
XDA & 0.896 & 0.789 & 0.839 \\

XDA(pruned)          & 0.972 & 0.921 & 0.946 \\
\bottomrule
\end{tabular}
\end{table} 
Our manual investigation of this obfuscated binary revealed a comparatively straightforward obfuscation methodology: the basic blocks of the protected function were scattered to distant locations in memory. Apart from this scattering, the instructions themselves appeared standard. However, IDA and Ghidra failed to correctly identify the function's entry point, resulting in exceptionally low recall. In contrast, DeepDi, ddisasm, and our \textit{TadyA} model, which utilize superset disassembly incorporating global graph information (due to control flow edges between these spatially distant instructions), successfully recalled them. \textit{Tady} missed two instructions, but successfully recalled them after pruning. XDA, despite achieving higher recall than IDA Pro, was still constrained by its reliance on local sequential context. However, applying our post-processing algorithm to XDA effectively propagated information along the execution trace, boosting its recall from $0.789$ to $0.921$. This serves as a clear example of how our post-processing algorithm can enhance model performance.

\section{Generalizability to Unseen Settings}
To further evaluate our models' ability to generalize to unseen settings, we conducted additional experiments. While their ability to handle unseen binaries can be inferred from the preceding experiments, here we specifically investigate generalization to unseen compilers and optimization levels. We explicitly splitted the Pangine dataset (used for training in previous experiments) by compiler and optimization level. We then conducted cross-validation to assess whether a model trained under one specific setting (e.g., a particular compiler and optimization level) could generalize to others. Specifically, we selected two compilers (clang-6.0.0 and gcc-7.5.0) and four optimization levels (O0-O3). For each combination, we trained the \textit{Tady} model for five epochs on the corresponding dataset partition. The increased epoch count was chosen to compensate for the smaller size of these individual training datasets. Detailed results are omitted for brevity, as the models generally generalized well across settings, achieving F1 scores close to 1.0 when tested on configurations different from their training set. The only notable exception was the model trained on clang-6.0.0 with O0 optimization; while its precision remained near 1.0, its recall on other settings was slightly lower, but still achieved around 0.98. These results demonstrate \textit{Tady}'s robust generalization capabilities.
 \end{document}